\documentclass[11pt]{article}
\title{Non-Existence of Time-Periodic Solutions of the Dirac Equation
in a Reissner-Nordstr\"om Black Hole Background}
\author{Felix Finster\thanks{Research supported by the Schweizerischer 
Nationalfonds.},\ 
Joel Smoller\thanks{Research supported in part by the NSF, Grant No.\ 
DMS-G-9501128.}, and Shing-Tung Yau\thanks{Research supported in part 
by the NSF, Grant No.\ 33-585-7510-2-30.}}
\date{May 1998}

\newtheorem{Def}{Def.}[section]
\newtheorem{Thm}[Def]{Theorem}

\newtheorem{Lemma}[Def]{Lemma}
\newtheorem{Remark}[Def]{Remark}
\newcommand{\Proof}{{\em{Proof: }}}
\newcommand{\QED}{\ \hfill $\FBox$ \\[1em]}

\newcommand{\spc}{\;\;\;\;\;\;\;\;\;\;}
\newcommand{\bra}{\mbox{$< \!\!$ \nolinebreak}}
\newcommand{\ket}{\mbox{\nolinebreak $>$}}

\newcommand{\1}{\mbox{\rm 1 \hspace{-1.05 em} 1}}

\newcommand{\sZ}{\mbox{\rm \bf \scriptsize Z}}
\newcommand{\sR}{\mbox{\rm \scriptsize I \hspace{-.8 em} R}}

\newcommand{\FBox}{\rule{2mm}{2.25mm}}

\include{epsf}

\begin{document}
\maketitle

\begin{abstract}
It is shown analytically that the Dirac equation has no normalizable, time-periodic
solutions in a Reissner-Nordstr\"om black hole background; in 
particular, there are no static solutions of the Dirac equation in 
such a background metric. The physical interpretation is that Dirac particles 
can either disappear into the black hole or escape to infinity, but they
cannot stay on a periodic orbit around the black hole.
\end{abstract}

\section{Introduction}
In recent years, there has been much interest in the gravitational collapse of matter
to a black hole. Although both analytical \cite{Christo} and intensive numerical
studies (see e.g.\ \cite{Numerik}) have given some understanding of how the event
horizon and the singularity form, little is known about the asymptotic form of the
black hole as $t \rightarrow \infty$. This is mainly due to the fact that standard
numerical methods become unreliable after the solutions have formed singularities.
Since all matter on a microscopic level is formed out of Dirac particles, it seems
especially interesting to study the asymptotic collapse of a ``cloud'' of
spin-$\frac{1}{2}$-particles. As a first step towards this goal, in this 
paper we study Dirac particles in a Reissner-Nordstr\"om background field.

We remark  that  considerable work has been done in the study of
quantum mechanical wave equations in the presence of black holes. The
papers which are most related are \cite{N1, N2}, where a massless Dirac
particle is considered in a Schwarzschild metric background, and asymptotic
completeness  is shown for the scattering states near the event horizon
and at infinity. However, the  most physically interesting case of a massive
Dirac particle near a charged black hole has not yet been considered. As we
will see here, both the rest mass of the Dirac particle and the charge
of the black hole lead to interesting physical effects and require new
analytical tools.

In polar coordinates $(t,r, \vartheta, \varphi)$, the Reissner-Nordstr\"om metric has
the form
\begin{equation}
        ds^2 \;=\;  \left(1 - \frac{2 \rho}{r} \:+\: 
        \frac{q^2}{r^2}\right) dt^2 \:-\: \left(1 - \frac{2 \rho}{r} \:+\: 
        \frac{q^2}{r^2}\right)^{-1} dr^2 \:-\: r^2 
        \:(d\vartheta^2 + \sin^2 \vartheta \:d\varphi^2) \;\;\; ,
        \label{eq:30}
\end{equation}
where $q$ is the charge of the black hole and $\rho$ its 
(ADM) mass. Furthermore, we have an external
electromagnetic potential $A$ of the form $A=(-\phi, \vec{0})$ with the
Coulomb potential
\begin{equation}
\phi(r) \;=\; \frac{q}{r} \;\;\; .
        \label{eq:cp}
\end{equation}
If $q<\rho$, the metric has two horizons; this is the so-called
{\em{non-extreme case}}. If $q=\rho$, the metric has only one horizon at $r=\rho$; this {\em{extreme case}} describes a black hole at zero temperature; cf.\ 
\cite{Hawking, Wald, SW}. For $q>\rho$, the metric does 
not describe a black hole, and thus this case will not be considered.

We describe the Dirac particles with ``classical'' wave functions (i.e.\ without
second quantization). Both the gravitational and electric fields are coupled to the
Dirac particles. {\em{We do not assume any spatial symmetry on the wave functions.}}
Near a collapsing black hole, it seems reasonable that that some of the Dirac
particles could get into static or time-periodic states. Our main 
result is to show that this is not possible.

In the following we will restrict to time-periodic solutions, noting 
that static solutions are a special case.
For classical point particles, the time-periodic solutions describe closed orbits
of particles rotating around the black hole. Our goal is to investigate how this
classical picture changes by the introduction of relativistic wave mechanics
and spin.
Since the phase of the Dirac wave function $\Psi$ is of no physical significance, we
say that $\Psi$ is {\em{periodic with period $T$}} if
\begin{equation}
        \Psi(t+T, r, \vartheta, \varphi) \;=\; e^{-i \Omega T} 
        \:\Psi(t,r,\vartheta, \varphi)
        \label{period}
\end{equation}
for some real $\Omega$. Our main result in the non-extreme case is the
following theorem:
\begin{Thm}
\label{thm1}
In a non-extreme Reissner-Nordstr\"om black-hole background, there 
are no normalizable, periodic solutions of the Dirac equation.
\end{Thm}
In the extreme case, we prove a slightly weaker statement:
\begin{Thm}
\label{thm2}
In an extreme Reissner-Nordstr\"om background, every normalizable, time-periodic
solution of the Dirac equation vanishes identically for $r>\rho$.
\end{Thm}
This surprising result shows that the classical picture breaks down completely; for
Dirac particles, there are no periodic solutions. This means that Dirac particles
which are attracted by a Reissner-Nordstr\"om black hole either ``fall into'' the
singularity or escape to infinity, but they cannot stay on a periodic orbit around the
black hole. The result can also be applied to the Dirac particles of 
the matter in the gravitational collapse; it then indicates that all the matter must
eventually disappear in the black hole.

Basically, our result is a consequence of the Heisenberg Uncertainty Principle and of
the particular form of the Dirac current. As a preparatory step, we
analyze the behavior of the Dirac wave functions near the event horizon 
and we derive conditions which relate the wave function outside and inside 
the horizon. It is essential for our methods and results that the particles have
spin. This shows that the spin is an important effect to be taken
into account in the study of gravitational collapse.

In the remainder of this section, we give some basic formulas needed 
to describe Dirac particles in curved space-time (for a more detailed introduction
to the classical Dirac theory in curved space-time see \cite{F}).
In this paper, the Dirac equation is always of the form
\begin{equation}
        \left( i G^j(x) \:\frac{\partial}{\partial x^j} \:+\:\frac{i}{2} 
        \:(\nabla_j G^j)(x)  \:+\: e \:G^j(x) \:A_j(x) \right) \Psi(x)
        \;=\; m \: \Psi(x) \;\;\; ,
        \label{eq:0}
\end{equation}
where $m$ is the rest mass of the particle, $A=A_j dx^j$ the electromagnetic 
potential, and $e$ the electromagnetic coupling constant (see 
\cite{FSY}, \cite{FSY2} for a derivation of this equation).
The Dirac matrices $G^j(x)$ are real linear combinations of the usual
$\gamma$-matrices. We work in the Dirac representation
\begin{equation}
        \gamma^0 \;=\; \left( \begin{array}{cc} \1 & 0 \\ 0 & -\1 
        \end{array} \right) \;\;\;,\spc \gamma^i \;=\;
        \left( \begin{array}{cc} 0 & \sigma^i \\ -\sigma^i & 0
        \end{array} \right) \;\;,\;\;\; i=1,2,3,
        \label{2.3}
\end{equation}
where $\sigma^i$ denote the Pauli matrices.
The Dirac matrices are related to the Lorentzian metric via the 
anti-commutation relations
\begin{equation}
        g^{jk}(x) \;=\; \frac{1}{2} \left\{ G^j(x),\:G^k(x) \right\} 
        \;\;\; .
        \label{eq:2a}
\end{equation}
The term $\nabla_j G^j$ in (\ref{eq:0}) is the divergence with respect to
the Levi-Civita connection; it can be easily computed via the standard formula
\begin{equation}
        \nabla_j G^j \;=\; \frac{1}{\sqrt{|g|}} \:\partial_j (\sqrt{|g|} 
        \:G^j) \;\;\; .
        \label{eq:01}
\end{equation}
For the normalization of the wave functions, one takes a space-like hypersurface
${\cal{H}}$ with normal vector field $\nu$ and considers the scalar product
\begin{equation}
        (\Psi \:|\: \Phi) \;=\; \int_{\cal{H}} \overline{\Psi} G^j \Phi 
        \:\nu_j \:d\mu \;\;\; ,
        \label{eq:f}
\end{equation}
where $\overline{\Psi}=\Psi^* \gamma^0$ is the adjoint spinor, and 
where $d\mu$ is the invariant measure on ${\cal{H}}$ induced by the 
Lorentzian metric. On solutions of the Dirac equation, we impose the
normalization condition
\[ (\Psi \:|\: \Psi)=1 \;\;\; . \]
Current conservation
\begin{equation}
        \nabla_j \:\overline{\Psi} G^j \Psi \;=\; 0
        \label{15a}
\end{equation}
implies that this normalization condition remains unchanged if the 
hypersurface ${\cal{H}}$ is continuously deformed.

\section{The Dirac Operator in a Schwarzschild Background}
\label{sec2}
\setcounter{equation}{0}
We begin by analyzing the Dirac operator in a Schwarzschild 
background metric. Our aim is to analyze the behavior of 
the spinors near the event horizon. To do this, we must consider the Dirac 
equation in different coordinate systems.

\subsection{The Dirac Operator in Polar Coordinates}
\label{sec21}
In polar coordinates $(t, r, \vartheta, \varphi)$, the Schwarzschild 
metric is
\[ ds^2 \;=\; \left(1-\frac{2\rho}{r} \right) dt^2 \:-\: 
\left(1-\frac{2\rho}{r} \right)^{-1} dr^2 \:-\: r^2 \:(d \vartheta^2 + \sin^2 
\vartheta \:d\varphi^2) \;\;\; , \]
where $\rho$ is the (ADM) mass. The metric has an event horizon at 
$r=2\rho$. In order to derive the Dirac operator, we first choose 
Dirac matrices $G^j(x)$ satisfying the anti-commutation relations (\ref{eq:2a}). 
The Dirac operator is then obtained by calculating the divergence 
(\ref{eq:01}) and substituting into (\ref{eq:0})\footnote{We point out that 
the choice of the Dirac matrices is not canonical; there are 
different real linear combinations of the $\gamma$-matrices which 
satisfy (\ref{eq:2a}). But the Dirac operators 
corresponding to different choices of the Dirac matrices are
equivalent in the sense that they can be obtained from 
each other by a 
suitable local transformation of the spinors (see e.g.\ \cite{F}). For 
this reason, we can simply choose the $G^j$ in the way which is most 
convenient to us.}.

Outside the horizon, we can satisfy the anti-commutation relations
(\ref{eq:2a}) by choosing the Dirac matrices in the form
\begin{equation}
        G^t = \frac{1}{S} \:\gamma^t \;\;,\;\;\;\; G^r = S 
        \:\gamma^r \;\;,\;\;\;\; G^\vartheta = \gamma^\vartheta 
        \;\;,\;\;\;\; G^\varphi = \gamma^\varphi \spc (r>2\rho)
        \label{eq:d}
\end{equation}
with
\[ S(r) \;=\; \left| 1 - \frac{2\rho}{r} \right|^{\frac{1}{2}} \;\;\; , \]
where $\gamma^t$, $\gamma^r$, $\gamma^\vartheta$, and $\gamma^\varphi$ 
are the ``$\gamma$-matrices in polar coordinates''
\begin{eqnarray}
\gamma^t &=& \gamma^0 \nonumber \\
\gamma^r &=& \gamma^3 \:\cos \vartheta \:+\: \gamma^1 \:\sin 
\vartheta \: \cos \varphi \:+\: \gamma^2 \:\sin \vartheta \:\sin \varphi 
\label{eq:p1} \\
\gamma^\vartheta &=& \frac{1}{r} \left(-\gamma^3 \:\sin \vartheta \:+\:
\gamma^1 \:\cos \vartheta \: \cos \varphi \:+\: \gamma^2 \:\cos \vartheta
\:\sin \varphi \right) \\
\gamma^\varphi &=& \frac{1}{r\:\sin \vartheta} \left(
-\gamma^1 \:\sin \varphi \:+\: \gamma^2 \:\cos \varphi \right) \;\;\; .
\label{eq:p2}
\end{eqnarray}
The divergence of the Dirac matrices is computed to be
\[ \nabla_j G^j \;=\; \left(S^\prime \:+\: \frac{2}{r} \:(S-1)
\right) \gamma^r \;\;\; . \]
Substituting into (\ref{eq:0}), we obtain for the Dirac operator,
$G_{\mbox{\scriptsize{out}}}$, in the region $r>2\rho$
\begin{equation}
        G_{\mbox{\scriptsize{out}}}
        \;=\; \frac{i}{S} \:\gamma^t \:\frac{\partial}{\partial t} \:+\: 
        \gamma^r \left(i S\: \frac{\partial}{\partial r} \:+\: 
        \frac{i}{r} \:(S-1) \:+\: \frac{i}{2} \:S^\prime \right) 
        \:+\: i \gamma^\vartheta \frac{\partial}{\partial \vartheta}
        \:+\: i \gamma^\varphi \frac{\partial}{\partial \varphi} \;\;\; .
        \label{eq:1}
\end{equation}
For the normalization, we integrate over the hypersurface 
$t={\mbox{const}}$; i.e.
\begin{equation}
        ( \Psi \:|\: \Psi)_{\mbox{\scriptsize{out}}}^t \;:=\; \int_{\sR^3 
        \setminus B_{2\rho}} (\overline{\Psi} \gamma^t \Psi)(t, \vec{x})
        \;S^{-1}\; d^3 x \;\;\; ,
        \label{eq:2}
\end{equation}
where $B_{2\rho}$ denotes the ball of radius $2\rho$ around the origin.
This normalization integral is problematic near the event horizon, as
will be discussed in detail later.
Inside the horizon, we must take into account that the radial 
direction $r$ is time-like, whereas $t$ is a space coordinate. So in this 
region, to obtain the Dirac matrices, we reverse the roles of the matrices
$\gamma^t$ and $\gamma^r$,
\begin{equation}
        G^t = \frac{1}{S} \:\gamma^r \;\;,\;\;\;\; G^r = -S 
        \:\gamma^t \;\;,\;\;\;\; G^\vartheta = \gamma^\vartheta 
        \;\;,\;\;\;\; G^\varphi = \gamma^\varphi \spc (r<2\rho).
        \label{eq:e}
\end{equation}
The divergence of the Dirac matrices now has the form
\[ \nabla_j G^j \;=\; -\frac{2}{r} \:\gamma^r \:-\: 
\left(S^\prime \:+\: \frac{2}{r}\:S \right) \gamma^t \;\;\; . \]
Thus the Dirac operator, $G_{\mbox{\scriptsize{in}}}$, in the region $r<2\rho$ is
given by
\begin{equation}
        G_{\mbox{\scriptsize{in}}}
        \;=\; \gamma^r \left( \frac{i}{S} \:\frac{\partial}{\partial t} 
        \:-\: \frac{i}{r} \right) \:-\: \gamma^t \left( i S 
        \:\frac{\partial}{\partial r} \:+\: \frac{i}{r} \:S \:+\: 
        \frac{i}{2} \: S^\prime \right)
        \:+\: i \gamma^\vartheta \frac{\partial}{\partial \vartheta}
        \:+\: i \gamma^\varphi \frac{\partial}{\partial \varphi} \;\;\; .
        \label{eq:3}
\end{equation}
According to (\ref{eq:f}), the naive extension of the normalization integral
(\ref{eq:2}) to the interior of the horizon is
\begin{equation}
        ( \Psi \:|\: \Psi)_{\mbox{\scriptsize{in}}}^t \;:=\;
        \int_{B_{2\rho}} (\overline{\Psi} \gamma^r \Psi)(t, \vec{x}) \;S^{-1}\;
        d^3 x \;\;\; ; \label{eq:4}
\end{equation}
this will also be discussed in detail later.

Notice that as a particular freedom in the choice of the Dirac matrices, the signs 
in (\ref{eq:d}) and (\ref{eq:e}) are arbitrary. As remarked above, 
this arbitrariness can be compensated by a suitable local transformation 
of the spinors.
However, this transformation of the spinors may change the sign 
of the scalar product (\ref{eq:f}). This is a subtle point which needs 
some explanation. Assume that we consider the space-like hypersurface outside
the horizon
\begin{equation}
        {\cal{H}}_1=\{t={\mbox{const}},\:r>2 \rho\} \;\;\; .
        \label{eq:h1}
\end{equation}
Its normal vector field $\nu$ is only determined up to a sign. 
Depending on whether we choose for $\nu$ the future or past directed normals,
the corresponding scalar product (\ref{eq:f}) will (for a fixed choice of the
Dirac matrices $G^j$) be either positive or negative (semi-)definite.
However, the overall sign of the scalar product is of no
physical relevance; e.g., we could just redefine (\ref{eq:f}) by inserting a
minus sign. In order to fix the sign convention, we will in 
the following always assume that the scalar product (\ref{eq:f}) is positive
for the future-directed normal vector field (this convention is consistent 
with our choices (\ref{eq:d}) and (\ref{eq:2})).
The situation becomes more interesting if we also look 
at the region inside the horizon. For this, we consider 
the ``cylindric'' space-like hypersurface
\begin{equation}
        {\cal{H}}_2 = \{r = r_0, \; t_0 \leq t \leq t_1 \}
        \label{eq:h2}
\end{equation}
for some fixed $r_0<2 \rho$ and $t_0<t_1$. A short
computation shows that, for our choice of the Dirac 
matrices (\ref{eq:e}), the scalar product (\ref{eq:f}) corresponding 
to ${\cal{H}}_2$ is positive if we choose for $\nu$ the inner normal 
(pointing towards the singularity at $r=0$). According to our sign 
convention, this means that the inward radial direction points to the 
future. Thus the particles ``fall into'' the singularity as time 
progresses, and we have a {\em{black hole}}.
On the other hand, we could have chosen the Dirac matrices such that the 
scalar product corresponding to ${\cal{H}}_2$ is positive for the 
outer normal (e.g.\ by changing the sign of $G^r$ in (\ref{eq:e})). In this case, 
increasing $r$ would correspond to going forward in time, and we would have
a {\em{white hole}}. Notice that this argument is consistent with time
reversals. Namely, the replacement $t \rightarrow -t$ forces us to change the 
sign of the scalar product (\ref{eq:f}) (in order that (\ref{eq:f}) is still
positive for ${\cal{H}}={\cal{H}}_1$ and future-directed normals). As a
consequence, the scalar product corresponding to ${\cal{H}}_2$ changes sign.
This means that black holes become white holes and vice versa.
We conclude that the Dirac operators $G_{\mbox{\scriptsize{out}}}$ and
$G_{\mbox{\scriptsize{in}}}$ distinguish between a black and a
white hole. This is a peculiar effect of the Dirac operator. It is 
quite different from e.g.\ the wave operator describing scalar fields 
(the Klein-Gordon operator), which does not determine the direction of time
inside the horizon.

Our description of the spinors in polar coordinates is not quite
satisfactory. First of all, the normalization integral inside the 
horizon, (\ref{eq:4}), is not definite.
This is a consequence of the fact that the $t$-variable is space-like 
inside the horizon. From the mathematical point of view, this is no 
problem; it seems tempting to just integrate across the horizon by 
adding (\ref{eq:2}) and (\ref{eq:4}). On the other hand, it is a
conceptual difficulty that the integrand in (\ref{eq:4}) is not 
positive and therefore does not have the
interpretation as a probability density. Furthermore, the Dirac 
equations corresponding to $G_{\mbox{\scriptsize{out}}}$ and
$G_{\mbox{\scriptsize{in}}}$ separately describe the wave functions 
outside and inside the horizon. But it is not clear how to match the wave functions
on the horizon.
For a better understanding of these issues, it is useful to remove
the singularity of the metric on the horizon by transforming to Kruskal
coordinates.

\subsection{Kruskal Coordinates}
\begin{figure}[tb]
        \centerline{\epsfbox{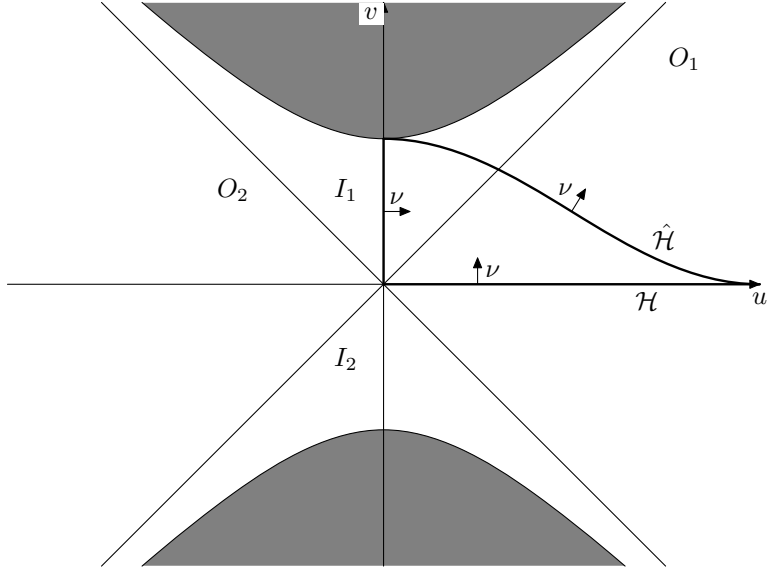}}
        \caption{Kruskal Coordinates}
        \label{fig1}
\end{figure}

According to \cite{ABS}, we introduce Kruskal coordinates $u$ and $v$ by
\begin{eqnarray}
u & = & \left\{ \begin{array}{ll}
\displaystyle \sqrt{\frac{r}{2\rho}-1} \:e^{\frac{r}{4\rho}} \:\cosh \left(
\frac{t}{4\rho}\right) & {\mbox{for $r>2\rho$}} \\[.8em]
\displaystyle \sqrt{1-\frac{r}{2\rho}} \:e^{\frac{r}{4\rho}} \:\sinh \left(
\frac{t}{4\rho}\right) & {\mbox{for $r<2\rho$}} \end{array} \right. 
\label{eq:11c} \\[.3em]
v & = & \left\{ \begin{array}{ll}
\displaystyle \sqrt{\frac{r}{2\rho}-1} \:e^{\frac{r}{4\rho}} \:\sinh \left(
\frac{t}{4\rho}\right) & {\mbox{for $r>2\rho$}} \\[.8em]
\displaystyle \sqrt{1-\frac{r}{2\rho}} \:e^{\frac{r}{4\rho}} \:\cosh \left(
\frac{t}{4\rho}\right) & {\mbox{for $r<2\rho$}}  \;\;\; . \end{array} \right.
\label{eq:11d}
\end{eqnarray}
The regions $r>2\rho$ outside and $r<2\rho$ inside the horizon 
are mapped into
\[ O_1 \;=\; \{u>0,\: |v|<u \} \]
and
\[ I_1 \;=\; \{ v>0,\: |u|<v ,\: v^2-u^2<1\} \;\;\; , \]
respectively (see Figure \ref{fig1}).
The horizon $r=2\rho$ corresponds to the 
origin $u=0=v$, and the lines $v=\pm u$ are reached in the limit $t 
\rightarrow \pm \infty$. Finally, the singularity at $r=0$ corresponds to
the hyperbola $v^2-u^2=1$, $v>0$.

In Kruskal coordinates $(v,u, \vartheta, \varphi)$, the Schwarzschild 
metric takes the form
\begin{eqnarray*}
ds^2 &=& f^{-2} \:(dv^2 - du^2) \:-\: r^2(d\vartheta^2 + \sin^2 
\vartheta \: d\varphi^2) \spc {\mbox{with}} \\
f^{-2} &=& \frac{32 \rho^3}{r} \: e^{-\frac{r}{2\rho}} \;\;\; .
\end{eqnarray*}
This metric is regular except at the singularity $v^2-u^2=1$; it can 
be extended to the entire region $v^2-u^2 < 1$.

Since the metric is regular at the origin, we can smoothly 
extend the Dirac operator across the horizon. To do this, we simply view
$v$ and $u$ as the time and space variables, respectively. We choose for the
Dirac matrices
\[ G^v = f \:\gamma^t \;\;,\;\;\;\; G^u = f
\:\gamma^r \;\;,\;\;\;\; G^\vartheta = \gamma^\vartheta 
\;\;,\;\;\;\; G^\varphi = \gamma^\varphi \;\;\; . \]
A straightforward computation yields for the Dirac operator
\begin{eqnarray}
G &=& \gamma^t \left( f i\:\frac{\partial}{\partial v} \:+\: \frac{i}{r} \:
f \:(\partial_v r)\:-\: \frac{i}{2} \:\partial_v f \right)
\:+\: \gamma^r \left( fi\:\frac{\partial}{\partial u} \:+\: \frac{i}{r}
\:(f \:(\partial_u r)-1) \:-\:\frac{i}{2} \:\partial_u f \right) \nonumber \\
&&+\: i \gamma^\vartheta \partial_\vartheta
                + i \gamma^\varphi \partial_\varphi \;\;\; .
        \label{eq:5}
\end{eqnarray}
The normalization integrals (\ref{eq:2}) and (\ref{eq:4}) on the 
surface $t=0$ correspond in Kruskal coordinates to the integral
(\ref{eq:f}) with
\[ {\cal{H}} \;=\; \{u=0,\:0 \leq v \leq 1\} \cup \{v=0,\:u>0 \} \;\;\; . \]
We choose the normal $\nu$ as in Figure \ref{fig1}. Using the current
conservation (\ref{15a}), one
can continuously deform the hypersurface ${\cal{H}}$ without changing the value 
of the normalization integral. 
In particular, we can avoid integrating across the horizon by choosing 
the hypersurface $\hat{{\cal{H}}}$ in Figure~\ref{fig1}. This is a major 
advantage of Kruskal coordinates; it gives a physically reasonable 
positive normalization integral even inside the event horizon. However, this
method must be done with care when the considered solution of the Dirac
equation has singularities near the origin.
Unfortunately, our time-periodic solutions of the Dirac equation will, after
transforming to Kruskal coordinates, in general be highly singular at
the origin. Therefore the deformation of the hypersurface as in
Figure~\ref{fig1} would be problematic, and we will not use this method.
In order to avoid any difficulties of the normalization integral near
the horizon, {\em{we shall only consider the normalization integral outside
and away from the event horizon}}.

\subsection{Transformation of the Dirac Operator}
We now  consider how the Dirac operator (\ref{eq:1}), (\ref{eq:3}) in 
polar coordinates transforms into the Dirac operator (\ref{eq:5}) in 
Kruskal coordinates. This transformation consists of
transforming both the space-time coordinates and the spinors. For 
clarity, we perform these transformations in two separate steps. Under 
the transformation of the space-time coordinates, the partial 
derivatives transform as
\begin{eqnarray*}
\frac{\partial}{\partial t} &=& \frac{\partial v}{\partial t} 
\:\frac{\partial}{\partial v} \:+\: \frac{\partial u}{\partial t} 
\:\frac{\partial}{\partial u} \;=\; \frac{1}{4\rho} \left( u 
\:\frac{\partial}{\partial v} \:+\: v\: \frac{\partial}{\partial u} 
\right) \\
\frac{\partial}{\partial r} &=& \frac{\partial v}{\partial r} 
\:\frac{\partial}{\partial v} \:+\: \frac{\partial u}{\partial r} 
\:\frac{\partial}{\partial u}
\;=\; \left\{ \begin{array}{ll}
        \displaystyle \frac{1}{4 \rho \:S^2} \left( v \:\frac{\partial}{\partial 
        v} \:+\: u \:\frac{\partial}{\partial u} \right) \spc& {\mbox{for 
        $r>2\rho$}} \\[.8em]
        \displaystyle -\frac{1}{4 \rho \:S^2} \left( v \:\frac{\partial}{\partial 
        v} \:+\: u \:\frac{\partial}{\partial u} \right) & {\mbox{for 
        $r<2\rho \;\;\; .$}}
\end{array} \right.
\end{eqnarray*}
Substituting into (\ref{eq:1}) and (\ref{eq:3}) gives for the Dirac 
operators $G_{\mbox{\scriptsize{out}}}$ and 
$G_{\mbox{\scriptsize{in}}}$ in Kruskal coordinates
\begin{eqnarray}
G_{\mbox{\scriptsize{out}}} &=& \frac{i}{4 \rho S} \:(u 
\gamma^t + v \gamma^r) \:\frac{\partial}{\partial v} \:+\: 
\frac{i}{4 \rho S} \:(v \gamma^t + u \gamma^r) \:\frac{\partial}{\partial 
u} \nonumber \\
&&+ \left( \frac{i}{r}\:(S-1) \:+\:\frac{i}{2} \:S^\prime 
\right) \gamma^r \:+\: i \gamma^\vartheta \:\frac{\partial}{\partial 
\vartheta} \:+\: i \gamma^\varphi \:\frac{\partial}{\partial 
\varphi} \label{eq:12a} \\
G_{\mbox{\scriptsize{in}}} &=& \frac{i}{4\rho S} \:(v 
\gamma^t + u \gamma^r) \:\frac{\partial}{\partial v} \:+\: 
\frac{i}{4 \rho S} \:(u \gamma^t + v \gamma^r) \:\frac{\partial}{\partial 
u} \nonumber \\
&&- \left( \frac{i}{r}\:S \:+\:\frac{i}{2} \:S^\prime 
\right) \gamma^t \:-\:\frac{i}{r} \:\gamma^r\:+\: i \gamma^\vartheta \:\frac{\partial}{\partial 
\vartheta} \:+\: i \gamma^\varphi \:\frac{\partial}{\partial \varphi} \;\;\; .
\label{eq:12b}
\end{eqnarray}
These Dirac operators do not coincide with (\ref{eq:5}), and we must therefore 
perform a further transformation; namely a transformation of the 
spinors. Under general coordinate transformations, the wave functions 
transform according to
\begin{equation}
\Psi(x) \:\rightarrow\: U(x) \:\Psi(x) \;\;\; ;
        \label{eq:28}
\end{equation}
in the case considered here, $U$ is the time-dependent $(4 \times 4)$ matrix
\begin{equation}
U(t) \;=\; \cosh \left(\frac{t}{8 \rho}\right) \:\1 \:+\: \sinh 
\left(\frac{t}{8 \rho}\right) \:\gamma^t \:\gamma^r \;\;\; .
        \label{eq:10}
\end{equation}
Under this transformation, the Dirac operators
$G_{\mbox{\scriptsize{out/in}}}$ transform as
\begin{equation}
        G_{\mbox{\scriptsize{out/in}}} \;\rightarrow\; U
        G_{\mbox{\scriptsize{out/in}}} U^{-1} \;\;\; .
        \label{eq:29}
\end{equation}
This gives the Dirac operator (\ref{eq:5}) in Kruskal coordinates,
\[ G \;=\; U \:G_{\mbox{\scriptsize{out}}}\:U^{-1} \;=\;
U \:G_{\mbox{\scriptsize{in}}}\:U^{-1} \;\;\; , \]
and this can be verified as follows: Under the transformation 
(\ref{eq:29}), the Dirac matrices behave like
\[ G^j(x) \;\rightarrow\; U(x) \:G^j(x) \:U(x)^{-1} \;\;\; . \]
Using the form of the Dirac matrices in (\ref{eq:12a}), (\ref{eq:12b})
and the explicit formula (\ref{eq:10}), a short calculation shows 
that the Dirac matrices of the operators $U G_{\mbox{\scriptsize{out/in}}} 
U^{-1}$ coincide with the Dirac matrices $f \gamma^t$, $f \gamma^r$, 
$\gamma^\vartheta$, and $\gamma^\varphi$ in (\ref{eq:5}). According 
to (\ref{eq:0}), the Dirac operator in the gravitational field is formed 
from the Dirac matrices and their covariant derivatives; it is thus 
completely determined by the Dirac matrices. Therefore, the operators
$U G_{\mbox{\scriptsize{out/in}}} U^{-1}$ must coincide with $G$. (One can 
also verify explicitly that the zero order terms of the operators
$U G_{\mbox{\scriptsize{out/in}}} U^{-1}$ and $G$ are
equal. This is a longer computation, however.)
We conclude that the Dirac operators $G_{\mbox{\scriptsize{out}}}$ 
and $G_{\mbox{\scriptsize{in}}}$ can be identified with the Dirac 
operator $G$ in the region $O_1 \cup I_1 = \{u+v>0,\: v^2-u^2<1\}$.

We remark that it is not possible to map the interior of the horizon into
the region
\[ I_2 \;=\; \{v<0, \:|u|<-v,\: v^2 - u^2<1\} \]
and still match the Dirac operator
$G_{\mbox{\scriptsize{in}}}$ with $G$, because this would contradict 
the fact discussed in the previous section that the Dirac operator 
distinguishes between black and white holes.

Finally, we note that the transformation (\ref{eq:28}),(\ref{eq:29}) of the
spinors can be viewed as a local $U(2,2)$ gauge transformation; see \cite{F}.

\subsection{Matching of the Spinors on the Horizon}
\label{sec24}
We now come to the question of how the wave functions inside and outside 
the horizon are related to each other. For this, we analyze the behavior of
solutions of the Dirac equation at the origin in Kruskal coordinates. After
transforming back to polar coordinates, this will give matching conditions
for the wave functions on the event horizon. The physical situation 
which we have in mind is a Dirac particle attracted by a 
Schwarzschild black hole. It suffices to do the matching for static solutions,
(and not time-periodic solutions), since in Section~\ref{sec4},
we reduce the problem to static solutions.

Let $\Psi$ be a static wave function, i.e.\ in polar coordinates
\[ \Psi(t,r,\vartheta,\varphi) \;=\; e^{-i \omega t} \: \Psi(r,\vartheta,\varphi)
\;\;\; . \]
We assume that $\Psi(r, \vartheta, \varphi)$ is a smooth function both 
inside and outside the horizon; i.e.\ in the regions $r<2 \rho$ and
$r>2 \rho$; this will be justified later by a separation of
variables technique. Furthermore, we assume that $\Psi$
is a solution of the Dirac equations
$(G_{\mbox{\scriptsize{in}}}-m) \Psi = 0$ and $(G_{\mbox{\scriptsize{out}}}-m) \Psi =
0$, respectively. According to the transformation rules (\ref{eq:28}),(\ref{eq:10}),
the wave function $\Psi$ in Kruskal coordinates takes the form
\begin{equation}
\Psi(u,v, \vartheta, \varphi) \;=\; U(t) \;e^{-i \omega t} \:\Psi(r, 
\vartheta, \varphi) \;\;\; ,
        \label{eq:7a}
\end{equation}
where $r$ and $t$ are given implicitly in terms of $u$ and $v$ by
\begin{eqnarray}
\left(\frac{r}{2 \rho}-1\right) e^{\frac{r}{2 \rho}} & = & u^2 - v^2
\label{eq:11e} \\
\tanh \left(\frac{t}{4 \rho} \right) & = & \left\{ \begin{array}{ll}
        \displaystyle \frac{v}{u} & {\mbox{for $r > 2 \rho$}} \\[0.8em]
        \displaystyle \frac{u}{v} & {\mbox{for $r < 2 \rho \;\;\; .$}}
\end{array} \right.
\end{eqnarray}
With this procedure, $\Psi$ is only defined in $O_1 \cup I_1$, the upper right
half of the Kruskal domain; it solves the Dirac equation
\begin{equation}
(G-m)\:\Psi \;=\; 0 \label{eq:30u}
\end{equation}
in the open set $O_1 \cup I_1$. If $\Psi$ is to be a physically
reasonable solution of the Dirac equation, it 
must be possible to extend it to the entire region $v^2-u^2<1$ between 
the two hyperbolas. If 
this extended wave function was not zero in the region $u+v<0$, our 
system would be connected to a white hole or to another universe 
(through a worm hole), and the Dirac particle would have a certain 
probability to be in these extensions of space-time. Since we are 
only interested in black holes, this is not the situation we 
want to consider. Therefore we demand that the extension of $\Psi$ must 
vanish identically in the half plane $u+v<0$.
We conclude that in Kruskal coordinates, we must analyze a solution 
$\Psi$ of the Dirac equation (\ref{eq:30u}) of the form
\begin{equation}
\Psi(u,v,\vartheta, \varphi) \;=\; \left\{ \begin{array}{cl}
U(t) \: e^{-i \omega t} \:\Psi(r, \vartheta, \varphi)
& {\mbox{for $u+v>0$, $u \neq v$}} \\ 0
& {\mbox{for $u+v<0 \;\;\; .$}}
\end{array} \right. \label{eq:18c}
\end{equation}
This wave function may be singular on the lines $u=\pm v$; in this 
case, $\Psi$ must solve the Dirac equation in a generalized weak
sense.

For the calculation of the weak derivatives of $\Psi$, we rewrite the wave
function in the form
\[ \Psi \;=\; \Theta(u+v) \:\Theta(u-v) \:\Psi_O \:+\: \Theta(v+u) \: \Theta(v-u) \:
\Psi_I \;\;\; , \]
where $\Psi_O=\Psi_{|O_1}$ and $\Psi_I=\Psi_{|I_1}$ are the components of $\Psi$
outside resp.\ inside the horizon ($\Theta$ denotes the Heaviside 
function $\Theta(x)=1$ for $x \geq 0$ and $\Theta(x)=0$ otherwise).
Since $\Psi$ satisfies the Dirac equation in $O_1
\cup I_1$, we need only consider the singular contributions on the lines $u=\pm v$.
A formal calculation gives
\begin{eqnarray}
0 &=& (G-m) \:\Psi \nonumber \\
&\stackrel{(\ref{eq:5})}{=}& f i \:\left(\gamma^t + \gamma^r) \:\delta(u+v) \;
(\Theta(u-v) \:\Psi_O \:+\: \Theta(v-u) \:\Psi_I \right)
\label{eq:u1} \\
&&-f i \:(\gamma^t - \gamma^r) \:\delta(u-v) \:\Theta(u+v) \:(\Psi_O - \Psi_I) \;\;\; .
\label{eq:u2}
\end{eqnarray}
If $\Psi_O$ and $\Psi_I$ were smooth up to the boundary of $O_1$ resp.\ $I_1$,
this equation would be well-defined in the distributional sense.
In general, however, $\Psi_O$ and $\Psi_I$ might be singular in the limit
$u \rightarrow \pm v$. In order to treat this
general case, we multiply (\ref{eq:u1}),(\ref{eq:u2}) with test functions 
$\eta(u,v)$ which, as $u \rightarrow \pm v$, decay so fast that the integral
over the resulting expression is well-defined. Since the matrices 
$(\gamma^t+\gamma^r)$ and $(\gamma^t-\gamma^r)$ are linearly 
independent, we get the two conditions
\begin{eqnarray}
\int_{\sR^2} \eta \:f\: \delta(u+v) \: (\gamma^t + \gamma^r) \:\left(
\Theta(u-v) \:\Psi_O \:+\: \Theta(v-u) \:\Psi_I \right) du\:dv
\label{eq:u3} &=& 0 \\
\int_{\sR^2} \eta \:f\: \delta(u-v) \: (\gamma^t - \gamma^r) \:\Theta(u+v)
\:(\Psi_O - \Psi_I) \:du\:dv &=& 0 \;\;\; . \label{eq:u4}
\end{eqnarray}
In (\ref{eq:u3}), we compensate the possible divergence of $\Psi$ for
$u \rightarrow -v$ by choosing $\eta$ in the region $O_1 \cup I_1$ to 
be of the form $\eta_{|O_1 \cup I_1} = (1+|(\gamma^t + \gamma^r) \Psi
|)^{-1}\:g$ with a smooth function $g$. Then the integrand in (\ref{eq:u3}) is of the
form $\delta(u+v) \times g \times (\mbox{bounded function})$, and the integral makes sense.
Since $g$ is arbitrary, we conclude that the integrand on the line $u=-v$ must
vanish, which implies that
\begin{equation}
\lim_{u \rightarrow -v} \:(\gamma^t + \gamma^r) \:\Psi(u,v,\vartheta,\varphi) \;=\; 0
\;\;\; . \label{eq:u5}
\end{equation}
In (\ref{eq:u4}), we can proceed similarly; namely, if $\Psi$ is singular on
the line $u=v$, we compensate the divergence of the integrand in (\ref{eq:u4})
by choosing $\eta$ to have an appropriately fast decay near the line $u=v$.
One must keep in mind, however, that $\eta$ cannot be chosen 
independently in $O_1$ and $I_1$, because the smoothness of $\eta$ on 
the line $u=v$ may impose restrictions on $\eta$. For example, if $\Psi_O$ and 
$\Psi_I$ have poles near $u=v$,
\begin{eqnarray*}
\Psi_I(u,u+\varepsilon, \vartheta, \varphi) &=& c_1(u,\vartheta, 
\varphi) \:\varepsilon^{-p} \:+\: \cdots \\
\Psi_O(u,u-\varepsilon, \vartheta, \varphi) &=& c_2(u,\vartheta, 
\varphi) \:\varepsilon^{-q} \:+\: \cdots\spc (\varepsilon>0),
\end{eqnarray*}
then we must choose $\eta$ in the form
\[ \eta(u,u+\varepsilon, \vartheta, \varphi) \;=\; 
c_3(u,\vartheta,\varphi) \:\varepsilon^{\max(p,q)} \:+\: \cdots \spc 
(\varepsilon>0 {\mbox{ or }} \varepsilon<0). \]
Thus the asymptotic behavior of $\eta$ near $u=v$ in $O_1$ and $I_1$ must be
the same. In the integral (\ref{eq:u4}), this means that the leading 
order singularities of $\Psi_O$ and $\Psi_I$ may cancel each other 
for any choice of $\eta$. Therefore, the condition for the leading 
order singularity takes the form
\begin{eqnarray}
\lefteqn{ (\gamma^t - \gamma^r) \left( \Psi(u,u+\varepsilon, \vartheta, \varphi) -
\Psi(u,u-\varepsilon, \vartheta, \varphi) \right) } \nonumber \\
&=& o(1+|(\gamma^t - \gamma^r) \:\Psi(u,u+\varepsilon,
\vartheta, \varphi)|) \spc {\mbox{as $\varepsilon \rightarrow 0$.}}
\label{eq:u6}
\end{eqnarray}
If the singularity of $\Psi$ on the line $u=v$ is worse than 
polynomial or of different form, there may be no obstructions for the 
choice of $\eta$ in $O_1$ and $I_1$. In this case, (\ref{eq:u6}) will 
still be a necessary condition. It will no longer be the strongest 
possible condition, but this is irrelevant for our purposes. For 
simplicity, we will use (\ref{eq:u6}) in the general case.

Next we evaluate the conditions (\ref{eq:u5}) and (\ref{eq:u6}) for our
wave function $\Psi(u,v,\vartheta,\varphi)$ in (\ref{eq:18c}). Using
(\ref{eq:10}), we have in $O_1 \cup I_1$
\begin{eqnarray*}
(\gamma^t + \gamma^r) \:\Psi(u,v,\vartheta,\varphi) &=& (\gamma^t + \gamma^r) \:
e^{\frac{t}{8 \pi}} \:e^{-i \omega t} \:\Psi(r,\vartheta,\varphi) \\
(\gamma^t - \gamma^r) \:\Psi(u,v,\vartheta,\varphi) &=& (\gamma^t - \gamma^r) \:
e^{-\frac{t}{8 \pi}} \:e^{-i \omega t} \:\Psi(r,\vartheta,\varphi) \;\;\; .
\end{eqnarray*}
The explicit formulas (\ref{eq:11c}) and (\ref{eq:11d}) enable us to write the
time-exponential in terms of $u$ and $v$ as
\[ e^{\pm \frac{t}{8 \rho}} \;=\; \left| \frac{u+v}{u-v} \right|^{\pm \frac{1}{4}}
\;\;\; . \]
Using the relation (\ref{eq:11e}) between $r$, $u$, and $v$, the condition
(\ref{eq:u5}) in polar coordinates takes the form
\begin{equation}
\lim_{\varepsilon \rightarrow 0} (\gamma^t + 
\gamma^r) \:|\varepsilon|^{\frac{1}{4}}
\Psi(t, 2\rho + \varepsilon, \vartheta, \varphi) \;=\; 0 \;\;\; .
\label{eq:mc1}
\end{equation}
Similarly, condition (\ref{eq:u6}) can be written in polar coordinates as
\begin{eqnarray}
\lefteqn{ (\gamma^t - \gamma^r) \:|\varepsilon|^{\frac{1}{4}} \:\left(
\Psi(2 \rho + \varepsilon,\vartheta, \varphi) \:-\:
\Psi(2 \rho - \varepsilon, \vartheta, \varphi) \right) } \nonumber \\
&=& o(1+|\varepsilon|^{\frac{1}{4}} \:|(\gamma^t - \gamma^r) \:\Psi(2 \rho +
\varepsilon, \vartheta, \varphi)|) \spc {\mbox{as $\varepsilon \rightarrow 0$.}}
\label{eq:u8}
\end{eqnarray}
In order to simplify this formula, we consider the decomposition of $\Psi$ in the form
\begin{equation}
|\varepsilon|^{\frac{1}{4}} \:\Psi \;=\; \frac{1}{2} \:\gamma^t \left(
(\gamma^t + \gamma^r) \:|\varepsilon|^{\frac{1}{4}} \:\Psi \:+\:
(\gamma^t - \gamma^r) \:|\varepsilon|^{\frac{1}{4}} \:\Psi \right) \;\;\; .
\label{eq:u7}
\end{equation}
According to condition (\ref{eq:mc1}), the first summand in the bracket
in (\ref{eq:u7}) vanishes on
the horizon $r=2 \rho$. Since the matrix $\gamma^t$ is invertible, we conclude that
$\Psi$ and $(\gamma^t - \gamma^r) \Psi$ are of the same order on the horizon.
Thus we can take out the matrices $(\gamma^t - \gamma^r)$ in (\ref{eq:u8}) and finally
obtain the equivalent condition
\begin{equation}
|\varepsilon|^{\frac{1}{4}} \left(
\Psi(t,2 \rho + \varepsilon, \vartheta, \varphi) \:-\: \Psi(t,2 \rho -
\varepsilon, \vartheta, \varphi) \right) \;=\; o(1
+|\varepsilon|^{\frac{1}{4}} \:\Psi(t,2 \rho + \varepsilon, \vartheta, \varphi))
\label{eq:mc2}
\end{equation}
as $\varepsilon \rightarrow 0$. The relations (\ref{eq:mc1}) and (\ref{eq:mc2})
are our {\em{matching conditions}}.

We briefly explain what these matching conditions mean, without being 
mathematically precise. First of all, we point out that the matrix
$(\gamma^t + \gamma^r)$ in the first matching condition (\ref{eq:mc1})
is not invertible. Therefore, (\ref{eq:mc1}) does not imply that
$|\varepsilon|^{\frac{1}{4}} \:\Psi(2 \rho + \varepsilon, \vartheta, \varphi)$
goes to zero in the limit $\varepsilon \rightarrow 0$; in general, this limit 
need not even exist.
Although the matching conditions have a quite special form, they can be
understood intuitively if one considers the Dirac current in polar coordinates.
We first look at the total normalization integral (\ref{eq:2})+(\ref{eq:4}):
\begin{eqnarray*}
\lefteqn{ (\Psi \:|\: \Psi)_{\mbox{\scriptsize{out}}} +
(\Psi \:|\: \Psi)_{\mbox{\scriptsize{in}}} } \\
&=& \int_{B_{2 \rho}} 
\overline{\Psi} (\gamma^r + \gamma^t) \Psi \:d\mu \:-\: 
\int_{B_{2 \rho}} \overline{\Psi} \gamma^t \Psi \:d\mu \:+\;
\int_{\sR^3 \setminus B_{2 \rho}} \overline{\Psi} \gamma^t \Psi 
\:d\mu \;\;\; .
\end{eqnarray*}
The condition (\ref{eq:mc1}) ensures that the integral of the
first summand is small near the horizon. 
Using the matching condition (\ref{eq:mc2}), one sees that the 
integrals in the second and last summands behave similarly
near the horizon. Because of the opposite sign of the second and 
third summands, this tends to make the normalization integral
finite even if $\Psi$ is singular on the horizon (if the current had a 
pole, for example, one could define the normalization integral as a principal
value). Thus our matching conditions ``regularize'' the normalization integral across
the horizon.
It is also interesting to look at the current in radial direction. For this, we 
consider the normalization integral through the hypersurface ${\cal{H}}_2$,
(\ref{eq:h2}). For the outer normal $\nu$, this gives inside the horizon
\begin{equation}
(\Psi \:|\: \Psi)_{{\cal{H}}_2} \;=\;
-\int_{{\cal{H}}_2} \overline{\Psi} \gamma^t \Psi \:d\mu \spc (r<2 \rho).
\label{eq:20a}
\end{equation}
For $r>2 \rho$, on the other hand, we get the expression
\begin{eqnarray}
(\Psi \:|\: \Psi)_{{\cal{H}}_2} &=&
\int_{{\cal{H}}_2} \overline{\Psi} \gamma^r \Psi \:d\mu \nonumber \\
&=& \int_{{\cal{H}}_2} \overline{\Psi} (\gamma^r+\gamma^t) \Psi \:d\mu 
\:-\: \int_{{\cal{H}}_2} \overline{\Psi} \gamma^t \Psi \:d\mu \spc 
(r>2 \rho).
\label{eq:20b}
\end{eqnarray}
According to (\ref{eq:mc1}), the first integral in (\ref{eq:20b}) is small near the
horizon $r=2 \rho$. The matching condition (\ref{eq:mc2}) gives that the second
summand in (\ref{eq:20b}) behaves similar to (\ref{eq:20a}) near the horizon.
Thus our matching conditions tend to make the normalization integral through
${\cal{H}}_2$ a continuous function in $r_0$ on the horizon $r_0=2 \rho$. Since the
integrand of the normalization integral has the interpretation as the ``probability
density'' or ``probability current,'' this means physically that a particle
which disappears in the event horizon must reappear in the interior of the
horizon. This is in accordance with our physical assumption that there are no
other universes or white holes where the particle 
could disappear into or emerge from.

\section{Separation of the Angular and Time Dependence}
\setcounter{equation}{0}
We next study Dirac particles in the external 
Reissner-Nordstr\"om background fields (\ref{eq:30}), (\ref{eq:cp}).
Since the external fields are spherically symmetric and time-independent,
we can separate out the angular and time dependence of the 
wave functions via spherical harmonics and plane waves, respectively.
This is done in a manner similar to the central force problem in Minkowski
space (see e.g.\ \cite{S}).

We start with a compilation of some formulas involving the angular 
momentum operator $\vec{L}=-i (\vec{x} \times \vec{\nabla})$ (see 
e.g.\ \cite{LL}). Its square is
\[ L^2 \;=\; - \Delta_{S^2} \;=\; L_+ L_- + L_z^2 - L_z \;=\; L_- 
L_+ + L_z^2 + L_z \]
with $L_\pm = L_x \pm i L_y$.
The spherical harmonics $Y^k_l$, $l=0,1,\ldots$, $k=-l,\ldots,l$ are 
simultaneous eigenfunctions of $L^2$ and $L_z$, namely
\begin{equation}
        L^2 \:Y^k_l \;=\; l(l+1) \:Y^k_l \;\;\;,\spc L_z \:Y^k_l \;=\; k 
        \:Y^k_l \;\;\; .
        \label{eq:3a}
\end{equation}
They are orthonormal,
\[ \int_{S^2} Y^{k *}_l \:Y^{k^\prime}_{l^\prime} \;=\; \delta_{l 
l^\prime} \:\delta^{k k^\prime} \;\;\; , \]
and form a basis of $L^2(S^2)$. The operators $L_\pm$ serve as
``ladder operators'', in the sense that
\begin{equation}
        L_\pm \:Y^k_l \;=\; \sqrt{l(l+1) - k(k \pm 1)} \:Y^{k \pm 1}_l 
        \;\;\; .
        \label{eq:31}
\end{equation}
In preparation for the four-component Dirac spinors, we consider
two-component Pauli spinors. In analogy to (\ref{eq:p1})-(\ref{eq:p2}),
we denote the ``Pauli matrices in polar coordinates'' by
$\sigma^r$, $\sigma^\vartheta$, and $\sigma^\varphi$; i.e.,
\begin{eqnarray*}
\sigma^r &=& \sigma^3 \:\cos \vartheta \:+\: \sigma^1 \:\sin 
\vartheta \: \cos \varphi \:+\: \sigma^2 \:\sin \vartheta \:\sin \varphi \\
\sigma^\vartheta &=& \frac{1}{r} \left(-\sigma^3 \:\sin \vartheta \:+\:
\sigma^1 \:\cos \vartheta \: \cos \varphi \:+\: \sigma^2 \:\cos \vartheta
\:\sin \varphi \right) \\
\sigma^\varphi &=& \frac{1}{r\:\sin \vartheta} \left(
-\sigma^1 \:\sin \varphi \:+\: \sigma^2 \:\cos \varphi \right) \;\;\; .
\end{eqnarray*}
We have
\begin{eqnarray}
\sigma^\vartheta \partial_\vartheta + \sigma^\varphi 
\partial_\varphi & = & \vec{\sigma} \vec{\nabla} \:-\: \sigma^r 
\partial_r \;=\; \frac{\sigma^r}{r} \:(\vec{\sigma} \vec{x})
         (\vec{\sigma} \vec{\nabla} \:-\: \sigma^r \partial_r) \nonumber \\
&=& \frac{\sigma^r}{r} \left(r \partial_r \:+\:
         i \vec{\sigma} (\vec{x} \times \vec{\nabla}) \:-\: r \partial_r 
         \right) \;=\; -\frac{\sigma^r}{r} \:\vec{\sigma} \vec{L} \;\;\; ,
\label{eq:lx}
\end{eqnarray}
and thus
\begin{equation}
\vec{\sigma} \vec{L} \;=\; -r\:\sigma^r \:(\sigma^\vartheta 
\partial_\vartheta + \sigma^\varphi \partial_\varphi) \;\;\; .
        \label{eq:34}
\end{equation}

For $j=\frac{1}{2}, \frac{3}{2}, \ldots$ and $k=-j, -j+1, 
\ldots, j$, we introduce the two-spinors
\begin{eqnarray*}
\chi^k_{j-\frac{1}{2}} &=& \sqrt{\frac{j+k}{2j}} 
\:Y^{k-\frac{1}{2}}_{j-\frac{1}{2}} \: \left( 
\begin{array}{c} 1 \\ 0 \end{array} \right) \:+\:
\sqrt{\frac{j-k}{2j}} \:Y^{k+\frac{1}{2}}_{j-\frac{1}{2}} \: \left( 
\begin{array}{c} 0 \\ 1 \end{array} \right) \\
\chi^k_{j+\frac{1}{2}} &=& \sqrt{\frac{j+1-k}{2j+2}} 
\:Y^{k-\frac{1}{2}}_{j+\frac{1}{2}} \: \left( 
\begin{array}{c} 1 \\ 0 \end{array} \right) \:-\:
\sqrt{\frac{j+1+k}{2j+2}} \:Y^{k+\frac{1}{2}}_{j+\frac{1}{2}} \: \left( 
\begin{array}{c} 0 \\ 1 \end{array} \right) \;\;\; .
\end{eqnarray*}
These spinors form an orthonormal basis of $L^2(S^2)^2$. They are 
eigenvectors of the operator $K=\vec{\sigma} \vec{L}+1$. More 
precisely, (\ref{eq:3a}) and (\ref{eq:31}) imply that
\begin{eqnarray}
K \:\chi^k_{j-\frac{1}{2}} & = & \left( \begin{array}{cc}
L_z+1 & L_- \\ L_+ & -L_z+1 \end{array} \right) \:\chi^k_{j-\frac{1}{2}}
\;=\; (j+\frac{1}{2}) \:\chi^k_{j-\frac{1}{2}}
\label{eq:37} \\
K \:\chi^k_{j+\frac{1}{2}} & = & -(j+\frac{1}{2}) \:\chi^k_{j+\frac{1}{2}}
\;\;\; . \label{eq:38}
\end{eqnarray}
Furthermore, multiplication with $\sigma^r$ again gives an eigenvector
of $K$; namely
\begin{eqnarray*}
K \:\sigma^r \chi^k_{j-\frac{1}{2}} &\stackrel{(\ref{eq:34})}{=}&
\left( -r \:\sigma^r \:(\sigma^\vartheta \partial_\vartheta + \sigma^\varphi 
\partial_\varphi) \:+1 \right) \sigma^r \:\chi^k_{j-\frac{1}{2}} \\
&=& -\sigma^r \:\chi^k_{j-\frac{1}{2}} \:-\: r\:\sigma^r \:(\sigma^\vartheta 
\:\sigma^r\:\partial_\vartheta + \sigma^\varphi 
\:\sigma^r\:\partial_\varphi) \:\chi^k_{j-\frac{1}{2}} \\
&=& -\sigma^r\:\chi^k_{j-\frac{1}{2}} \:-\: \sigma^r \:(\vec{\sigma} \vec{L}) 
\:\chi^k_{j-\frac{1}{2}} \\
&=& -\sigma^r \:K\: \chi^k_{j-\frac{1}{2}}
\;=\; -(j+\frac{1}{2}) \:\sigma^r \chi^k_{j-\frac{1}{2}} \;\;\; .
\end{eqnarray*}
Taking into account the normalization factors, we obtain the simple formula
\begin{equation}
        \sigma^r \chi^k_{j-\frac{1}{2}} \;=\; \chi^k_{j+\frac{1}{2}} 
        \;\;\; .
        \label{eq:35}
\end{equation}
Finally, we choose for the Dirac wave functions the two ansatz'
\begin{eqnarray}
\Psi^+_{jk \:\omega} &=& e^{-i \omega t} \:\frac{S^{-\frac{1}{2}}}{r}\:
        \left( \begin{array}{c} \chi^k_{j-\frac{1}{2}} \:\Phi^+_{jk \omega \:1}(r) \\
i \chi^k_{j+\frac{1}{2}}\:\Phi^+_{jk \omega \:2}(r) \end{array} \right)
\label{eq:36} \\
\Psi^-_{jk \:\omega} &=& e^{-i \omega t} \:\frac{S^{-\frac{1}{2}}}{r}\:
        \left( \begin{array}{c} \chi^k_{j+\frac{1}{2}} \:\Phi^-_{jk \omega \:1}(r) \\
i \chi^k_{j-\frac{1}{2}}\:\Phi^-_{jk \omega \:2}(r) \end{array} 
\right)
\label{eq:36a}
\end{eqnarray}
with the two-spinors $\Phi^+_{jk \omega}$ and $\Phi^-_{jk \omega}$.
A general solution of the Dirac equation can be written as a linear 
combination of these wave functions (this is because one can obtain every
combination of spherical harmonics in the four spinor components).

In the regions where the $t$-variable is time-like, we choose the Dirac
matrices again in the form (\ref{eq:d}), whereby the function $S$ is now 
given by
\begin{equation}
        S(r) \;=\; \left| 1 - \frac{2 \rho}{r} \:+\: \frac{q^2}{r^2} 
        \right|^{\frac{1}{2}} \;\;\; .
        \label{eq:1y}
\end{equation}
According to (\ref{eq:0}), the formula for the Dirac operator is obtained
by inserting the Coulomb potential into (\ref{eq:1}),
\begin{eqnarray}
G &=& \gamma^t \left(\frac{i}{S} \:\frac{\partial}{\partial t}
- \frac{e}{S} \:\phi \right) \:+\: \gamma^r
        \left( i S \:\frac{\partial}{\partial r} \:+\: \frac{i}{r}
        \:(S-1)\:-\: \frac{i}{2} \:S^\prime \right) \nonumber \\
&&\:+\: i \gamma^\vartheta \frac{\partial}{\partial \vartheta}
        \:+\: i \gamma^\varphi \frac{\partial}{\partial \varphi}
        \;\;\; . \label{eq:8a}
\end{eqnarray}
The identity (\ref{eq:lx}) allows to rewrite the angular derivatives 
of the Dirac operator in terms of the operator $K$. If we substitute the ansatz' 
(\ref{eq:36}),(\ref{eq:36a}) into the Dirac equation and apply the 
relations (\ref{eq:37}), (\ref{eq:38}), and (\ref{eq:35}), we obtain the
two-component Dirac equations
\begin{eqnarray}
\lefteqn{ S \:\frac{d}{dr} \Phi^\pm_{jk \omega} } \nonumber \\
&=& \left[ \left( 
\begin{array}{cc} 0 & -1 \\ 1 & 0 \end{array} \right) (\omega - e 
\phi) \:\frac{1}{S} \:\pm\:\left( \begin{array}{cc} 1 & 0 \\ 0 & -1 \end{array}
\right) \frac{2j+1}{2r} \:-\: \left(
\begin{array}{cc} 0 & 1 \\ 1 & 0 \end{array} \right) m \right]
\Phi^\pm_{jk \omega} \;\;\; . \spc
        \label{eq:de1}
\end{eqnarray}
In the regions where the $t$-direction is space-like, we obtain the
generalization of (\ref{eq:3}) for the Dirac operator; namely
\begin{eqnarray}
G &=& \gamma^r \left(\frac{i}{S} \:\frac{\partial}{\partial t}
-\frac{i}{r} \:-\:\frac{e}{S} \:\phi \right) \:+\: \gamma^0
        \left( i S \:\partial_r \:+\: S \:\frac{i}{r}
        \:+\: \frac{i}{2} \:S^\prime \right) \nonumber \\
&&+\: i \gamma^\vartheta \partial_\vartheta
        + i \gamma^\varphi \partial_\varphi
        \;\;\; . \label{eq:8b}
\end{eqnarray}
We again choose the ansatz' (\ref{eq:36}),(\ref{eq:36a}).
This gives the two-component Dirac equations
\begin{eqnarray}
\lefteqn{ S \:\frac{d}{dr} \Phi^\pm_{jk \omega} } \nonumber \\
&=& \left[ \left( 
\begin{array}{cc} 0 & -1 \\ 1 & 0 \end{array} \right) (\omega - e 
\phi) \:\frac{1}{S} \:\pm\:i \left( \begin{array}{cc} 0 & 1 \\ 1 & 0
\end{array} \right) \frac{2j+1}{2r} \:+\:i \left(
\begin{array}{cc} 1 & 0 \\ 0 & -1 \end{array} \right) m \right]
\Phi^\pm_{jk \omega} \;\;\; . \spc
        \label{eq:de2}
\end{eqnarray}

\section{Non-Extreme Reissner-Nordstr\"om Background}
\setcounter{equation}{0}
\label{sec4}
In this section, we consider the case  $q \neq \rho$, so that the 
metric coefficient $S(r)$, (\ref{eq:1y}), has two zeros
\[ r_0 \;=\; \rho - \sqrt{\rho^2 - q^2} \spc {\mbox{and}} \spc
r_1 \;=\; \rho + \sqrt{\rho^2 - q^2} \;\;\; . \]
These zeros are transversal, $S^\prime(r_j) \neq 0$; in addition, the 
potential $\phi(r)$ is regular at $r=r_j$.
Since our matching conditions (\ref{eq:mc1}) and (\ref{eq:mc2}) for 
the Schwarzschild metric only depend on the local behavior of the 
external field around the horizon, they are also valid for the 
Reissner-Nordstr\"om horizons (for the inner horizon, we must
reverse the $r$-direction). We will show in this section 
that these matching conditions do not admit {\em{normalizable}}, 
time-periodic solutions of the Dirac equation. More precisely, we will show 
that for every (non-trivial) solution of the Dirac equation (\ref{eq:0}), the
normalization integral outside and away from the horizons,
\begin{equation}
        (\Psi \:|\: \Psi)^t_\infty \;:=\; \int_{\sR^3 \setminus B_{2r_1}} 
        \overline{\Psi} \gamma^t \Psi \:S^{-1}\: d^3 x \;\;\; ,
        \label{eq:9a}
\end{equation}
is infinite for some $t$. Notice that for a normalized wave function, 
the integral (\ref{eq:9a}) gives the probability that the particle 
lies outside the ball of radius $2r_1$, which must be smaller than 
one. Thus, if (\ref{eq:9a}) is inifinite, the wave function cannot be 
normalized.

Suppose that we have a periodic solution (\ref{period}) of the Dirac equation with 
period $T$. Expanding the periodic function $e^{i \Omega 
t} \Psi(t,r,\vartheta, \varphi)$ in a Fourier series gives the 
representation of $\Psi$ (as the Bloch wave)
\begin{equation}
        \Psi(t,r,\vartheta,\varphi) \;=\; e^{-i \Omega t} \:\sum_{n \in 
        \sZ} \Psi_n(r, \vartheta, \varphi) \;e^{-2 \pi i \:n \:\frac{t}{T}} 
        \;\;\; .
        \label{eq:bw}
\end{equation}
Decomposing the functions $\Psi_n$ in the basis 
(\ref{eq:36}), (\ref{eq:36a}), and substituting into (\ref{eq:bw}) 
gives
\begin{equation}
\Psi(t,r,\vartheta,\varphi) \;=\; \sum_{n, j, k, s} \Psi^s_{j k 
\:\omega(n)}(t,r, \vartheta,\varphi) \;\;\; , \label{eq:43e}
\end{equation}
where the index $s=\pm$, and where $\omega$ is related to $n$ by
\[ \omega(n) \;=\; \Omega \:+\: \frac{2 \pi n}{T} \;\;\; . \]
Using the orthonormality of the two-spinors $\chi^k_{j \pm 
\frac{1}{2}}$, the normalization integral takes the form
\[ (\Psi \:|\: \Psi)^t_\infty \;=\; \int_{\sR^3 \setminus B_{2 r_1}}
\sum_{n, n^\prime} \sum_{j,k,s} \overline{\Psi^s_{j k\:\omega(n)}}
\gamma^t \Psi^s_{j k\:\omega(n^\prime)} \:S^{-1}\:d^3x \;\;\; . \]
The integrand has an oscillating time dependence of the form
$\exp(i(\omega(n)-\omega(n^\prime))t)$. In order to eliminate the
oscillations, we take the average over one period $(0,T)$, giving
\[ \frac{1}{T} \int_0^T (\Psi \:|\: \Psi)^t_\infty \:dt \;=\; 
\sum_{n, j, k, s} \:(\Psi^s_{j k \:\omega(n)} \:|\: \Psi^s_{j k 
\:\omega(n)})_\infty \;\;\; . \]
For a normalizable wave function $\Psi$, this expression is finite. 
Since the scalar product $(.|.)_\infty$ is (semi-)positive definite, we 
conclude that all the summands must be finite; thus
\begin{equation}
( \Psi^s_{j k \:\omega(n)} \:|\: \Psi^s_{j k \:\omega(n)}) \;<\; \infty
        \label{eq:8c}
\end{equation}
for all $s=\pm$, $j$, $k$, $n$.

This inequality allows us to turn our attention to the individual wave 
functions $\Psi^s_{jk\omega}$. As a first step we
show that the wave functions $\Phi^\pm$ in the ansatz' (\ref{eq:36}) 
and (\ref{eq:36a}) are not zero on the horizon.
\begin{Lemma}
\label{lemma1}
The function $|\Phi^\pm_{jk\omega}(r)|^2$ has finite boundary values on 
the horizon. If it is zero on a horizon $r=r_0$ or $r=r_1$, then
$\Phi^\pm_{jk\omega}$ vanishes identically.
\end{Lemma}
{\Proof}
For ease in notation, we omit the indices $j$, $k$, and $\omega$.
For a given $\delta$, $0<\delta<r_0$, the $t$-direction is time-like in the
regions $(\delta, r_0)$ and $(r_1, \infty)$. In these regions, the Dirac
equations (\ref{eq:de1}) give
\begin{eqnarray*}
S \:\frac{d}{dr} |\Phi^\pm|^2(r) &=& \bra S \:\frac{d}{dr} \Phi^\pm,\: \Phi^\pm \ket
\:+\: \bra \Phi^\pm,\: S \:\frac{d}{dr} \Phi^\pm \ket \\
&=& \pm\frac{2j+1}{r} 
\:(|\Phi^\pm_1|^2 - |\Phi^\pm_2|^2) \:-\: 4 m
\:{\mbox{Re}}\left((\Phi^\pm_1)^* \:\Phi^\pm_2 \right)
\end{eqnarray*}
and thus
\[ -c \:|\Phi^\pm|^2 \;\leq\; S \:\frac{d}{dr} |\Phi^\pm|^2 \;\leq\; 
        c \:|\Phi^\pm|^2 \]
with $c=2m+(2j+1)/\delta$. Dividing by $|\Phi^\pm|^2$ and integrating
yields, for $\delta < r < r^\prime < r_0$, or
$r_1 < r < r^\prime$, the inequality
\begin{equation}
        -c \int_r^{r^\prime} S^{-1} \;\leq\; \left. \log |\Phi^\pm|^2 
        \:\right|_r^{r^\prime} \;\leq\; c \int_r^{r^\prime} S^{-1} \;\;\; .
        \label{eq:i1}
\end{equation}
In the region $r_0<r<r_1$, the Dirac equations (\ref{eq:de2}) give similarly
\[ S \:\frac{d}{dr} |\Phi^\pm|^2(r)
\;=\; \bra S \:\frac{d}{dr} \Phi^\pm,\: \Phi^\pm \ket
\:+\: \bra \Phi^\pm,\: S \:\frac{d}{dr} \Phi^\pm \ket \;=\; 0 \;\;\; , \]
since the square bracket in (\ref{eq:de2}) is an anti-Hermitian matrix.
Thus $|\Phi^\pm|^2$ is constant in this region, and so, (\ref{eq:i1}) also 
(trivially) holds for $r_0<r<r^\prime<r_1$.

Notice that $S^{-1}$ is integrable on the event horizons. Therefore,
the inequality (\ref{eq:i1}) implies that the left and right sided boundary values of 
$|\Phi^\pm|^2$ on the horizon are finite, and are non-zero unless $\Phi^\pm$ 
vanishes identically in the corresponding region $(\delta, r_0)$, 
$(r_0, r_1)$, or $(r_1, \infty)$.

Next we consider the matching condition (\ref{eq:mc2}). If we substitute the ansatz'
(\ref{eq:36}) and (\ref{eq:36a}), we get for $\Phi^\pm$ the conditions
\[ \Phi^\pm(r_j+\varepsilon) - \Phi^\pm(r_j-\varepsilon) \;=\;
o(1+|\Phi^\pm(r_j+\varepsilon)|) \spc {\mbox{at $\varepsilon 
\rightarrow 0$}},j=0,1. \]
Since we have already shown that $|\Phi^\pm(r)|^2$ has two-sided 
limits as $r=r_j$, this last equality shows that the left and right sided
boundary values of $|\Phi^\pm|^2$ must coincide,
\[ \lim_{0<\varepsilon \rightarrow 0} |\Phi^\pm(r_j+\varepsilon)|^2 \;=\;
\lim_{0<\varepsilon \rightarrow 0} |\Phi^\pm(r_j-\varepsilon)|^2 \spc ,j=0,1. \]
We conclude that the wave function can only be zero 
on one of the horizons if it vanishes in the whole interval $(\delta, 
\infty)$. Taking the limit $\delta \rightarrow 0$ gives the result.
\QED
We point out that this lemma does not imply that the wave function $\Phi$
is continuous on the horizon. In general, $\Phi(r)$ will oscillate faster and 
faster as $r$ approaches a horizon. Nevertheless, its absolute value $|\Phi|$
tends to a finite value in this limit.

The next step is to use current conservation for analyzing the decay 
of $\Psi^s_{jk \:\omega(n)}$ at infinity.
\begin{Thm} {\bf{(radial flux argument)}}
Either $\Psi^s_{jk\omega}$ vanishes identically, or the normalization 
condition (\ref{eq:8c}) is violated.
\end{Thm}
{\Proof}
To simplify the notation, we again omit the indices $s$, $j$, $k$, and $\omega$.
Assume that $\Psi$ is not identically zero.
For $r_1<r<R$ and $T>0$, let $V=(0,T) \times (B_{2R} \setminus B_{2r})$ be 
an annulus outside the horizon. As a consequence of the current 
conservation, the flux integral over the boundary of $V$ is zero, thus
\begin{eqnarray*}
0 &=& \int_V \nabla_j (\overline{\Psi} G^j \Psi) \;\sqrt{|g|} \:d^4x \\
&=& \int_0^T dt \; r^2 \:S(r) \int_{S^2} (\overline{\Psi} \gamma^r 
\Psi)(t,r)
\:-\:\int_0^T dt \; R^2 \:S(R) \int_{S^2} (\overline{\Psi} \gamma^r 
\Psi)(t,R) \\
&&-\int_{2r}^{2R} ds \;s^2 \:S^{-1}(s) \int_{S^2} \left. 
(\overline{\Psi} \gamma^t \Psi)(t,r) \right|_{t=0}^{t=T} \;\;\; ,
\end{eqnarray*}
where $\int_{S^2}$ denotes the integral over the angular variables. 
Since the integrand is static, the last integral vanishes, and we obtain 
that the radial flux is independent of the radius,
\begin{equation}
         r^2 \:S(r) \int_{S^2} (\overline{\Psi} \gamma^r \Psi)(r) \;=\;
        R^2 \:S(R) \int_{S^2} (\overline{\Psi} \gamma^r \Psi)(R) \;\;\; .
        \label{eq:n1}
\end{equation}

We want to show that the radial flux is not zero. For this, we first 
substitute the ansatz' (\ref{eq:36}) and (\ref{eq:36a}) into the right 
side of (\ref{eq:n1}) and get
\begin{equation}
r^2 \:S(r) \int_{S^2} (\overline{\Psi} \gamma^r \Psi)(r) \;=\; 
\int_{S^2} \Phi^*(r) \left( \begin{array}{cc} 0 & i \\ -i & 0 
\end{array} \right) \Phi(r) \;\;\; .
        \label{eq:s1}
\end{equation}
According to Lemma \ref{lemma1}, $|\Phi|$ has finite, non-zero 
boundary values on the horizon $r_1$. Expressed in $\Phi$, the 
matching condition (\ref{eq:mc1}) gives
\[ \lim_{r_1 < r \rightarrow r_1} \left( \begin{array}{cc} 1 & i \\ i & 
-1 \end{array} \right) \Phi \;=\; 0 \;\;\; . \]
Using this equation, we take the limit $r \rightarrow r_1$ in 
(\ref{eq:s1}),
\begin{eqnarray*}
\lim_{r_1 < r \rightarrow r_1} r^2 \:S(r) \int_{S^2} (\overline{\Psi} \gamma^r \Psi)(r)
&=& \lim_{r_1 < r \rightarrow r_1} \int_{S^2} \left[ \Phi^* 
\left( \begin{array}{cc} 1 & i \\ -i & 1 \end{array} \right)
\Phi \:-\: |\Phi|^2 \right] \\
&=& \lim_{r_1 < r \rightarrow r_1} \int_{S^2} \left[ \Phi^* 
\left( \begin{array}{cc} 1 & 0 \\ 0 & -1 \end{array} \right)
\left( \begin{array}{cc} 1 & i \\ i & -1 \end{array} \right) \Phi
\:-\: |\Phi|^2 \right] \\
&=&-\lim_{r_1 < r \rightarrow r_1}
\int_{S^2} |\Phi|^2 \;\neq\; 0 \;\;\; ,
\end{eqnarray*}
where we used Lemma \ref{lemma1} in the last inequality

Now we consider the radial flux for large $R$. Since the flux is 
non-zero and independent of $R$, we have
\[ 0 \;<\; \lim_{R \rightarrow \infty} |R^2 \:S(R) \int_{S^2} 
(\overline{\Psi} \gamma^r \Psi)(R)| \;\;\; . \]
Using the positivity of the form $\overline{\Psi} \gamma^t \Psi$ and the 
fact that the Reissner-Nordstr\"om metric is asymptotically 
Minkowskian, we get (using the Cauchy-Schwarz inequality) the estimate
\begin{eqnarray*}
0 &<& \lim_{R \rightarrow \infty} |R^2 \:S(R) \int_{S^2} 
(\overline{\Psi} \gamma^r \Psi)(R)| \\
&\leq& \lim_{R \rightarrow \infty} |R^2 \:S(R) \int_{S^2} 
(\overline{\Psi} \gamma^t \Psi)(R)| \;=\;
\lim_{R \rightarrow \infty} |R^2 \:S^{-1}(R) \int_{S^2} 
(\overline{\Psi} \gamma^t \Psi)(R)| \;\;\; .
\end{eqnarray*}
We have shown that the integrand of our normalization integral
\[ (\Psi \:|\: \Psi)_\infty \;=\; \int_{2 r_1}^\infty dR \; R^2 
\:S^{-1}(R) \int_{S^2} (\overline{\Psi} \gamma^t \Psi)(R) \]
converges to a positive number. Thus the normalization integral 
must be infinite.
\QED
This theorem shows that the wave functions $\Psi^s_{jk \omega}$ in the decomposition
(\ref{eq:bw}), (\ref{eq:43e}) must all be identically zero. Thus there 
are no normalizable solutions of the Dirac equation; this proves Theorem \ref{thm1}.
\begin{Remark}
\em We point out that the radial flux argument is based only on our matching
conditions for the wave functions and on the Dirac current conservation. Therefore, it
can immediately be applied to more general static, spherically symmetric background
fields. This generalization may for example be relevant if the coupling of the
gravitational and electric field to matter or other force fields is taken into account. Although the
exact formulas of the Reissner-Nordstr\"om solution will then no longer be valid, the
qualitative behavior of the fields on the horizons may still be the same.
To give an example of the possible generalizations, we state the following theorem,
which can be proved with very similar methods:
\em Let $g_{ij}$ be a static, radially symmetric background metric,
\[ ds^2 \;=\; g_{ij} \:dx^i dx^j \;=\; \frac{1}{T^2(r)} \:dt^2 \:-\: \frac{1}{A(r)}
\:dr^2 \:-\: r^2 \:(d\vartheta^2 + \sin^2 \vartheta \:d\varphi^2) \;\;\; , \]
whereby the metric coefficient $A(r)$ has $N$ zeros at $r=r_1,\ldots,r_N$,
$0<r_0<\cdots<r_N$. Assume the following conditions hold:
\begin{description}
\item[(1)] The zeros of $A$ are all transversal,
\[ A'(r_j) \;\neq\; 0 \spc {\mbox{for $j=1,\ldots,N$}}. \]
\item[(2)] The determinant of the metric is regular except at the origin,
\[ T^{-2}(r) \:A^{-1}(r) \;\in\; C^\infty(0,\infty) \;\;\; . \]
\end{description}
Furthermore assume there is a spherically symmetric electric field $\phi(r)$ which is
regular except at the origin, $\phi \in C^\infty(0,\infty)$.
Then there are no normalizable, time-periodic solutions of the Dirac equation
with these background fields.
\end{Remark}

\section{Extreme Reiss\-ner-Nord\-str\"om Background}
\setcounter{equation}{0}
We now consider the case $q=\rho$ of an extreme Reissner-Nordstr\"om 
background field, i.e.
\[ S \;=\; \frac{r-\rho}{r} \;\;\; . \]
The metric coefficient $S$ now has only one zero at $r=\rho$; 
the $t$-direction is time-like both inside and outside the horizon.
This situation can be thought of as the limiting case that the two 
horizons $r_0$ and $r_1$ considered in the previous section come 
arbitrarily close. Unfortunately, the arguments for the non-existence 
proof do not carry over in this limit, so that we must rely on a different
method.

Since the $t$-direction is always time-like, the $t$-component of the 
current $\overline{\Psi} G^t \Psi$ is positive and has the usual 
interpretation as probability density. Therefore the normalization 
integral
\[ (\Psi \:|\: \Psi)^t \;=\; \int_{\sR^3} \overline{\Psi} 
\gamma^t \Psi \:S^{-1}\:d^3x \]
causes no conceptual difficulties.

Suppose that we had a normalizable, periodic solution (\ref{period})
of the Dirac equation with period $T$. Again using the representation as the 
Bloch wave (\ref{eq:bw}) and averaging over one period gives
\[ \infty \;>\; \frac{1}{T} \int_0^T (\Psi \:|\: \Psi)^t \:dt 
\;=\; \sum_{n, j, k, s} (\Psi^s_{j k \:\omega(n)} \:|\: \Psi^s_{j k 
\:\omega(n)}) \;\;\; . \]
Substituting the ansatz' (\ref{eq:36}) and (\ref{eq:36a}) yields
\[ \frac{1}{T} \int_0^T (\Psi \:|\: \Psi)^t \:dt
\;=\; \int_0^\infty dr \;S^{-2}(r) \:\sum_{n, j, k, s} |\Phi^s_{j k 
\:\omega(n)}|^2 \;\;\; . \]
Using the positivity of the summands, we obtain the conditions
\begin{equation}
\int_0^\infty dr \;S^{-2}(r) \: |\Phi^s_{j k \:\omega(n)}|^2 \;<\; 
\infty \label{eq:i0}
\end{equation}
for all $s$, $j$, $k$, and $n$.

We will now study the individual functions $\Phi^s_{j k \omega}$ for 
$r>\rho$. To simplify the notation, we again omit the indices
$j$, $k$, and $\omega$.
Our first task is to consider under which conditions on the parameters $\omega$, $j$,
and $m$ the normalization integral (\ref{eq:i0}) can be finite near $r=\rho$.
We first discuss the situation qualitatively:
Since $S^{-2}(r)=r^2/(r-\rho)^2$ has a non-integrable singularity on the 
horizon, the normalization integral will only be finite if $\Phi^s$ becomes small
near $r=\rho$. For generic parameter values, the dominant term in the Dirac equation
(\ref{eq:de1}) near $r=\rho$ is the first summand, i.e.
\[ \frac{d}{dr} \Phi^\pm \;\approx\; \frac{\omega-e\phi}{S^2}
\left( \begin{array}{cc} 0 & -1 \\ 1 & 0 \end{array} \right) \Phi^\pm \;\;\; . \]
Since in this limiting case, the eigenvalues of the matrix on the right are purely
imaginary, the Dirac equation describes fast oscillations of the wave function. The
eigenvalues of the second and third summands in (\ref{eq:de1}) are real; they
describe an exponential increase or decay of $\Phi$. If the oscillating term is
dominant, we expect that $\Phi$ will not go to zero in the
limit $r \rightarrow \rho$. In the following lemma, these ideas are made
mathematically precise in a slightly more general setting.
\begin{Lemma}
\label{lemma5}
Let $\Phi(x)$, $x>0$, be a nontrivial solution of the ODE
\begin{equation}
\Phi^\prime(x) \;=\; \left[ a(x) \:\left( \begin{array}{cc} 0 & -1 \\ 1 & 0
\end{array} \right) \:+\: b(x) \:\left( \begin{array}{cc} 1 & 0 \\ 0 & -1
\end{array} \right) \:+\: c(x) \:\left( \begin{array}{cc} 0 & 1 \\ 1 & 0
\end{array} \right) \right] \Phi(x) \label{eq:l0}
\end{equation}
with smooth, real functions $a, b, c \in C^\infty(0,\infty)$ and 
$a \neq 0$. If near the origin, the quotients $b/a$ and $c/a$ are monotone and
\begin{equation}
b(x)^2 + c(x)^2 \;<\; a(x)^2 \;\;\; , \label{eq:l5}
\end{equation}
then $|\Phi|^2(x)$ is bounded from above and from below near $x=0$,
\[ 0 \;<\; \liminf_{0<x \rightarrow 0} |\Phi(x)|^2 \;\leq\;
\limsup_{0<x \rightarrow 0} |\Phi(x)|^2 \;<\; \infty \;\;\; . \]
\end{Lemma}
{\Proof} Let $(0,\varepsilon)$ be an interval where the functions $b/a$ and $c/a$ are
monotone and where (\ref{eq:l5}) holds. Assume that $\Phi$ is a nontrivial solution of
(\ref{eq:l0}). According to the uniqueness theorem for the solutions of ODEs,
$\Phi(x)$ is non-zero for all $0<x<\infty$.
Now consider the functional
\[ F(x) \;=\; \bra \Phi(x), \:A(x) \:\Phi(x) \ket
\spc {\mbox{with}} \spc A(x) \;=\;
\left( \begin{array}{cc} 1+b/a & -c/a \\ -c/a & 1-b/a
\end{array} \right) \;\;\; . \]
According to (\ref{eq:l5}), the matrix $A$ is close to the identity; i.e., there is a
constant $c<1$ with
\[ |\1-A(x)| \;<\; c \spc {\mbox{for all $x$ with $0<x<\varepsilon$.}} \]
Thus the functional $F$ is uniformly bounded in $|\Phi|^2$ on 
$(0,\varepsilon)$,
\begin{equation}
\frac{1}{C} \:|\Phi(x)|^2 \;\leq\; F(x) \;\leq\; C \:|\Phi(x)|^2 \label{eq:l1}
\end{equation}
for some $C>0$. Using the special form of $A$ and of the differential equation
(\ref{eq:l0}), the derivative of $F$ takes the simple form
\begin{equation}
F^\prime(x) \;=\; \bra \Phi^\prime, \:A\:\Phi \ket \:+\:
\bra \Phi, \:A\:\Phi^\prime \ket \:+\: \bra \Phi, \:A^\prime\:\Phi \ket \;=\;
\bra \Phi, \:A^\prime\:\Phi \ket \;\;\; . \label{eq:l2}
\end{equation}
The sup-norm of the matrix $A^\prime$ is bounded by
\begin{equation}
|A^\prime| \;\leq\; \left| \left(\frac{b}{a}\right)^\prime \right| \:+\:
\left| \left(\frac{c}{a}\right)^\prime \right| \;\;\; . \label{eq:l3}
\end{equation}
Putting together (\ref{eq:l1}), (\ref{eq:l2}), and (\ref{eq:l3}), we get the bounds
\[ -C \:\left( \left| \left( \frac{b}{a} \right)^\prime \right| \:+\:
\left| \left( \frac{c}{a} \right)^\prime \right| \right) F(x) \;\leq\; F^\prime(x)
\;\leq\; C \:\left( \left| \left( \frac{b}{a} \right)^\prime \right| \:+\:
\left| \left( \frac{c}{a} \right)^\prime \right| \right) F(x) \;\;\; . \]
Now we divide by $F(x)$ and integrate. Since $b/a$ and $c/a$ are monotone, we can just
integrate inside the absolute values,
\begin{equation}
-C \:\left. \left( \left| \frac{b}{a} \right| \:+\: \left| \frac{c}{a} \right|
\right) \right|_x^y \;\leq\; \left. \log F \right|_x^y \;\leq\;
C \:\left. \left( \left| \frac{b}{a} \right| \:+\: \left| \frac{c}{a} \right|
\right) \right|_x^y \;\;\; .
\end{equation}
Since the extreme left and right sides of this inequality converge in the
limit $x \rightarrow 0$, we conclude that $\log F(x)$ is bounded from above and
below near the origin. After exponentiating and substituting (\ref{eq:l1}), the
result follows.
\QED
Applied to (\ref{eq:de1}), this lemma says that $|\Phi^\pm(r)|^2$ is
bounded away from zero near $r=\rho$ unless
\begin{equation}
\omega - e \:\phi(\rho) \;=\; 0 \;\;\; . \label{eq:u9}
\end{equation}
Thus we can turn our attention to this special case.

If we substitute the condition (\ref{eq:u9}) into (\ref{eq:de1}), 
the Dirac equation simplifies to
\begin{eqnarray}
\lefteqn{ \left|1-\frac{\rho}{r} \right| \:\frac{d}{dr} \Phi^\pm(r) }
\nonumber \\
&=& \left[ \left( 
\begin{array}{cc} 0 & -1 \\ 1 & 0 \end{array} \right) e \:\pm\:
        \left( \begin{array}{cc} 1 & 0 \\ 0 & -1 \end{array} \right)
        \frac{2j+1}{2r} \:-\: \left(
\begin{array}{cc} 0 & 1 \\ 1 & 0 \end{array} \right) m \right]
\Phi^\pm \;\;\; . \spc
        \label{eq:e0}
\end{eqnarray}
We want to study how the solutions of this equation behave for small $r-\rho>0$.
For this, we rewrite the equation in the new variable
\[ u(r) \;=\; -r - \rho \:\ln(r-\rho) \;\;\; , \]
which gives
\begin{equation}
\frac{d}{du} \Phi^\pm(u) \;=\; \left[ -\left( 
\begin{array}{cc} 0 & -1 \\ 1 & 0 \end{array} \right) e \:\mp\:
        \left( \begin{array}{cc} 1 & 0 \\ 0 & -1 \end{array} \right)
        \frac{2j+1}{2r} \:+\: \left( \begin{array}{cc} 0 & 1 \\ 1 & 0 \end{array}
        \right) m \right] \Phi^\pm \;\;\; . \label{eq:l10}
\end{equation}
The region near $r=\rho$ corresponds to large values of $u$. The matrix in the
bracket in (\ref{eq:l10}) depends smoothly on $u$ and converges in the limit $u
\rightarrow \infty$ to a finite limit , in view of the definition of $u$ given
above. According to the stable manifold theorem \cite[Thm.\ 4.1]{C}, the
solutions of (\ref{eq:l10}) which are not bounded away from zero for large $u$
tend exponentially to zero.
After transforming back to the variable $r$, this justifies the power ansatz
\begin{equation}
\Phi^\pm_1(r) \;=\; \Phi^\pm_{10} \;(r-\rho)^s \:+\: 
o((r-\rho)^s) \;\;\;,\;\;\;\;\;
\Phi^\pm_2(r) \;=\; \Phi^\pm_{20} \;(r-\rho)^s \:+\: 
o((r-\rho)^s)
        \label{eq:e1}
\end{equation}
with constants $\Phi^\pm_{10}$, $\Phi^\pm_{20}$ and a parameter $s>0$.
Substituting into (\ref{eq:e0}) yields the system of linear equations
\begin{eqnarray}
\left( s \mp (j+1/2) \right) \Phi^\pm_{10} &=& 
-\rho\:(m+e)\:\Phi^\pm_{20} \label{eq:e2} \\
\left( s \pm (j+1/2) \right) \Phi^\pm_{20} &=& 
-\rho\:(m-e)\:\Phi^\pm_{10} \;\;\; , \label{eq:e3}
\end{eqnarray}
which can be solved for $\Phi^\pm_{10}$ and $\Phi^\pm_{20}$.
In this way, we have found a consistent ansatz for the spinors near $r=\rho$. However,
the corresponding solutions of the Dirac equation are all not normalizable, as
the following theorem shows.
\begin{Thm}
Every nontrivial solution $\Phi^\pm(r)$, $r>\rho$, of the Dirac 
equation (\ref{eq:e0}) with the boundary conditions (\ref{eq:e1})
violates the normalization condition (\ref{eq:i0}).
\end{Thm}
{\Proof}
Let $\Phi^\pm$ be a nontrivial solution of the Dirac equation. Since 
the Dirac equation has real coefficients, we can
assume that $\Phi^\pm$ are real. In the new variable $u=r^{-1}$, the Dirac equation
(\ref{eq:e0}) takes the form
\begin{eqnarray*}
\lefteqn{ \left|1-\rho u \right| \:\frac{d}{du} \Phi^\pm(u) }
\nonumber \\
&=& \left[ -\frac{e}{u^2} \left( 
\begin{array}{cc} 0 & -1 \\ 1 & 0 \end{array} \right) \:\mp\:
        \frac{2j+1}{2u} \left( \begin{array}{cc} 1 & 0 \\ 0 & -1 \end{array} \right)
        \:+\: \frac{m}{u^2} \left(
\begin{array}{cc} 0 & 1 \\ 1 & 0 \end{array} \right) \right]
\Phi^\pm \;\;\; .
\end{eqnarray*}
If $e>m$, Lemma \ref{lemma5} yields that $|\Phi^\pm(u)|^2$ is bounded from above and
below near $u=0$. Thus $|\Phi^\pm(r)|^2$ does not decay at infinity, and the
normalization integral (\ref{eq:i0}) will diverge. We conclude that we must only
consider the case $m \geq e$.

In the case $m=e$, the system (\ref{eq:e2}), (\ref{eq:e3}) yields that either
$\Phi^\pm_{10}$ or $\Phi^\pm_{20}$ is zero. Furthermore, the Dirac equation
(\ref{eq:e0}) shows that either $\Phi^\pm_1$ or $\Phi^\pm_2$ vanishes
identically. Since $\Phi^\pm(r)$ has no zeros for finite $r$ (otherwise, the
uniqueness of the solution yields that $\Phi^\pm$ vanishes identically), we can
assume that the vector $\Phi^\pm(r)$ will lie in the fourth quadrant,
\begin{equation}
\Phi^\pm(r) \;\in\; \{ (x,y) \;|\; x \geq 0, \:y \leq 0\}
        \label{eq:e4}
\end{equation}
for all $r$.

Next we want to show that (\ref{eq:e4}) also holds in the case $m>e$.
In this case, from (\ref{eq:e2}) and (\ref{eq:e3}), we
can assume that $\Phi^\pm_{10}$ is positive, whereas $\Phi^\pm_{20}$ is negative.
Thus (\ref{eq:e4}) holds for small $r-\rho>0$. In order to show that the fourth
quadrant is an invariant region for $\Phi^\pm$, first notice 
that $\Phi^\pm(r)$ cannot become zero for a finite value of $r$. Thus, if
$\Phi^\pm(r)$
leaves the quadrant for some $r$, we have either
\[ \Phi^\pm_1(r)=0 \;\;\;,\spc(\Phi^\pm_1)^\prime(r) \leq 0 \;\;\;
{\mbox{and}}\spc \Phi^\pm_2(r) < 0 \]
or
\[ \Phi^\pm_1(r) > 0 \;\;\;,\spc
\Phi^\pm_2(r)=0\;\;\;{\mbox{and}}\spc (\Phi^\pm_2)^\prime(r) \geq 0 \;\;\; . \]
But the Dirac equation gives in the first case that $(\Phi^\pm_1)^\prime>0$ 
and in the second case that $(\Phi^\pm_2)^\prime(r)<0$, which is a 
contradiction.

\begin{figure}[tb]
        \centerline{\epsfbox{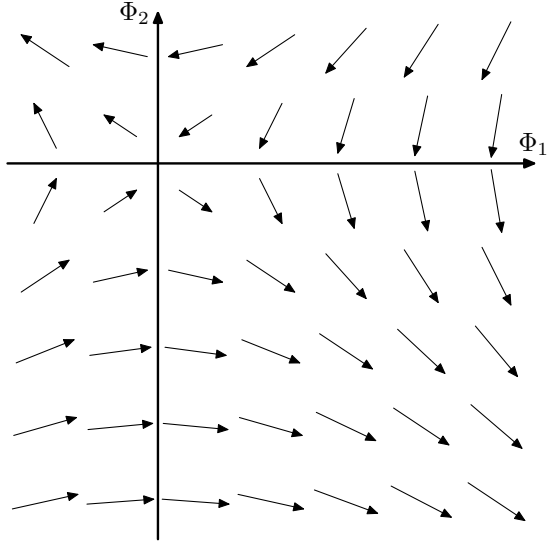}}
        \caption{Flow of $\Phi^\pm$ for large $r$, schematic}
        \label{fig2}
\end{figure}

We conclude that $\Phi^\pm(r)$ lies for all $r$ in the fourth quadrant. Figure \ref{fig2}
shows the flow of equation (\ref{eq:e0}) for large $r$.
From this one sees immediately that the origin is repelling, so that
$|\Phi^\pm|^2$ will be bounded away from zero for large $r$.
\QED
It follows that our periodic solution $\Psi$ must vanish identically outside
the horizon. This proves Theorem \ref{thm2}.

We point out that in contrast to the situation in Section \ref{sec4}, 
we do not make any statement on the behavior of the wave function 
for $r < \rho$. Indeed, it appears that the extreme Reissner-Nordstr\"om 
background does admit periodic solutions for $r<\rho$; these can be 
constructed by taking the boundary conditions (\ref{eq:e1}) on the 
horizon and solving the Dirac equation backwards in $r$.

\newpage
\appendix
\section{Justification of Time-Periodicity Inside the Horizon}
\setcounter{equation}{0}
Throughout this paper, we have considered a Dirac wave function (\ref{period})
which is time-periodic both inside and outside the event horizon.
Since an outside observer has no knowledge about the physical situation
in the interior of the event horizon, the assumption of time periodicity
inside the horizon might not seem physically resonable.
In this short appendix, we clarify why time periodicity inside the horizon
is natural to assume. Namely, we show that every solution $\Psi(t,r,\vartheta,
\varphi)$ of the Dirac equation which is time-periodic outside the event
horizon and (locally uniformly) bounded in $t$, gives rise to a solution
$\tilde{\Psi}$ of the Dirac equation, which coincides with $\Psi$
outside the horizon and is also time-periodic inside. Using this argument,
the results of this paper could be immediately generalized to Dirac
wave functions which are only time-periodic outside the event horizon.

Let $\Psi(t,r,\vartheta, \varphi)$ be a solution of the Dirac equation which
is time-periodic outside the event horizon,
\begin{equation}
\Psi(t+T, r, \vartheta, \varphi) \;=\; e^{-i \Omega T} \:\Psi(t,r,\vartheta,
\varphi) \;\;\;\;{\mbox{for $r>r_1$}}\;\;\;,
\label{A1}
\end{equation}
and locally uniformly bounded in $t$,
\begin{equation}
|\Psi(t,r,\vartheta,\varphi)| \;\leq\; F(r) \spc{\mbox{with}}\spc
F \in C^0((0,r_0) \cup (r_0, r_1))
\label{A2}
\end{equation}
($r_0$ and $r_1$ again denote the Cauchy and event horizons, 
respectively).
We consider for $N \geq 1$ the functions
\[ \tilde{\Psi}_N(t,r,\vartheta,\varphi) \;=\; \frac{1}{2N+1}
\sum_{n=-N}^N \Psi(t+nT, r, \vartheta, \varphi) \;\;\; . \]
Since our Dirac operator is static,
the functions $\tilde{\Psi}_N$ satisfy the Dirac equation.
Time-periodicity (\ref{A1}) implies that $\tilde{\Psi}_N$ and $\Psi$
coincide outside the event horizon. Inside the event horizon, one can use
the bound (\ref{A2}) to show that the $\tilde{\Psi}_N$ form a 
Cauchy sequence. Thus we can take the limit $N \rightarrow \infty$; we
set $\tilde{\Psi} = \lim_{N \rightarrow \infty} \tilde{\Psi}_N$. Again
using (\ref{A2}), we conclude that the function $\tilde{\Psi}$ is time
periodic,
\begin{eqnarray*}
\lefteqn{ \tilde{\Psi}(t+T, r, \vartheta, \varphi)
- \tilde{\Psi}(t, r, \vartheta, \varphi) } \\
&=& \lim_{N \rightarrow \infty} \frac{1}{2N+1} \left(\Psi(t+(N+1)T, r,
\vartheta, \varphi) - \Psi(t-NT, r, \vartheta, \varphi) \right) \;=\; 0 \;\;\;,
\end{eqnarray*}
and satisfies the Dirac equation,
\[ (G-m) \:\tilde{\Psi} \;=\; \lim_{N \rightarrow \infty} (G-m)
\:\tilde{\Psi}_N \;=\; 0 \;\;\; . \]

\begin{tabular}{ll}
\\
Mathematics Department, & Mathematics Department,\\
Harvard University, & The University of Michigan,\\
Cambridge, MA 02138  \hspace*{.5cm}(FF \& STY)\hspace*{1cm}
& Ann Arbor, MI 48109 \hspace*{.5cm} (JS)\\
\\
\end{tabular}

\begin{tabular}{ll}
email:&felix@math.harvard.edu\\
&smoller@umich.edu \\
&yau@math.harvard.edu
\end{tabular} 


\addcontentsline{toc}{section}{References}
\begin{thebibliography}{99}
\bibitem{Christo} Christodoulou, D., ``The formation of black holes 
and singularities in spherically symmetric gravitational collapse,'' 
{\em{Commun.\ Pure Appl.\ Math.}}\ 44 (1991) 339-373
\bibitem{Numerik} Choptuik, M.W., ``Universality and scaling in the 
gravitational collapse of a scalar field,'' {\em{Phys.\ Rev.\ Lett.}}\ 70
(1993) 9-12
\bibitem{N1}
Nicolas, J.-P., ``Scattering of linear {D}irac fields by a spherically
symmetric black hole,'' {\em{Ann.\ Inst.\ H.\ Poincar\'e}}, Physique
th\'eorique, 62 (1995) 145-179
\bibitem{N2}
Nicolas, J.-P., ``Op\'erateur de diffusion pour le syst\`eme de
{D}irac en m\'etrique de {S}chwarzschild,'' {\em{C.\ R.\ Acad.\ Sci.\ 
Paris}}\ (S\'erie I) 318 (1994) 729--734
\bibitem{Hawking} Hawking, S.W., ``Particle creation by black holes,''
{\em{Commun.\ Math.\ Phys.}}\ 43 (1975) 199-220
\bibitem{Wald} Wald, R., ``General Relativity,'' University of Chicago Press
(1984)
\bibitem{SW} Smoller, J., and A. Wasserman, ``Uniqueness of the extreme
Reissner-Nordstr\"om solution in $SU(2)$ Einstein-Yang-Mills theory for
spherically symmetric space-time,'' {\em{Phys.\ Rev.}}\ D, 52 (1995)
5812-5815
\bibitem{F} Finster, F., ``Local $U(2,2)$ symmetry in relativistic 
quantum mechanics,'' hep-th/9703083, {\em{J.\ Math.\ Phys.}}\ 39 (1998)
6276-6290
\bibitem{FSY} Finster, F., Smoller, J., and Yau, S.-T., 
``Particlelike solutions of the Einstein-Dirac equations,'' 
gr-qc/9801079, {\em{Phys.\ Rev.}}\ D 59 (1999) 104020
\bibitem{FSY2} Finster, F., Smoller, J., and Yau, S.-T., 
``Particlelike solutions of the Einstein-Dirac-Maxwell equations,'' 
gr-qc/9802012, {\em{Phys.\ Lett.}}\ A 259 (1999) 431-436
\bibitem{ABS} Adler, R., Bazin, M. and Schiffer, M., ``Introduction to 
General Relativity,'' 2nd edition, McGraw-Hill (1975)
\bibitem{S} Sakurai, J.J., ``Advanced Quantum Mechanics,'' Addison-Wesley, 1967
\bibitem{LL} Landau, L.D., Lifshitz, E.M., ``Quantum Mechanics,'' 
Pergamon Press (1977)
\bibitem{C} Coddington, E. and Levinson, N., ``Theory of Ordinary 
Differential Equations,'' McGraw-Hill (1955)
\end{thebibliography}
\end{document}